%% file: main.tex
\documentclass[11pt]{article}

\usepackage{bm, commath}
\usepackage{natbib}
\usepackage{caption}
\usepackage{graphicx}
\usepackage{subfig}
\usepackage{amsmath, amsfonts, amsthm}
\usepackage{float}
\usepackage{booktabs,siunitx}
\usepackage{url}
\usepackage{multirow}
\usepackage{amssymb}
\usepackage{blkarray}
\usepackage{bbm}
\usepackage{multicol}
\usepackage{authblk}
\usepackage{natbib}
\usepackage[shortlabels]{enumitem}
\usepackage{longtable}
\usepackage{adjustbox}
\usepackage{threeparttable}
\usepackage{hhline}
\usepackage{colortbl}
\usepackage{lscape}

\usepackage{listings}
\usepackage{bbm}
\usepackage{textcomp}
\usepackage[margin=1in]{geometry}
\usepackage{authblk}
\usepackage[ruled]{algorithm2e}
\SetKwInput{KwParam}{Parameter}
\SetAlgoCaptionLayout{centerline}

\usepackage{sectsty}
\usepackage[doublespacing]{setspace}
\usepackage[utf8]{inputenc}
\usepackage{float}
\usepackage{tikz}
\usetikzlibrary{shapes,decorations,arrows,calc,arrows.meta,fit,positioning}

\usepackage{mwe}
\usepackage{graphicx}

\usepackage[textsize=scriptsize, textwidth = 2cm, shadow]{todonotes}

\newtheorem{theorem}{Theorem}

\newtheorem{remark}{Remark}

\newtheorem*{assumption*}{Assumption}

\newtheorem{proposition}{Proposition}

\newtheorem{innercustomgeneric}{\customgenericname}
\providecommand{\customgenericname}{}
\newcommand{\newcustomtheorem}[2]{%
  \newenvironment{#1}[1]
  {%
   \renewcommand\customgenericname{#2}%
   \renewcommand\theinnercustomgeneric{##1}%
   \innercustomgeneric
  }
  {\endinnercustomgeneric}
}

\newcustomtheorem{customthm}{Theorem}
\newcustomtheorem{customassumption}{Assumption}

\usepackage{mathtools}

\usepackage{xcolor}

\makeatletter
\renewcommand{\algocf@captiontext}[2]{#1\algocf@typo. \AlCapFnt{}#2} 
\def\@algocf@capt@plain{top}
\renewcommand{\algocf@makecaption}[2]{%
  \addtolength{\hsize}{\algomargin}%
  \sbox\@tempboxa{\algocf@captiontext{#1}{#2}}%
  \ifdim\wd\@tempboxa >\hsize
  \hskip .5\algomargin%
  \parbox[t]{\hsize}{\algocf@captiontext{#1}{#2}}
  \else%
  \global\@minipagefalse%
  \hbox to\hsize{\box\@tempboxa}
  \fi%
  \addtolength{\hsize}{-\algomargin}%
}
\makeatother

\begin{document}

\sectionfont{\bfseries\large\sffamily}%

\subsectionfont{\bfseries\sffamily\normalsize}%





\title{Manipulating a Continuous Instrumental Variable in an Observational Study of Premature Babies:\\  
\Large Algorithm, Partial Identification Bounds, and Inference under Randomization and Biased Randomization Assumptions}

\author[1]{Zhe Chen $^\dagger$}
\author[2]{Min Haeng Cho $^\dagger$}
\author[3]{Bo Zhang\thanks{Assistant Professor of Biostatistics, Vaccine and Infectious Disease Division, Fred Hutchinson Cancer Center. Email: {\tt bzhang3@fredhutch.org}. $^\dagger$ The first two authors contributed equally. }}

\affil[1]{Department of Statistics, University of Illinois Urbana-Champaign}
\affil[2]{Department of Biostatistics, University of Washington}
\affil[3]{Vaccine and Infectious Disease Division, Fred Hutchinson Cancer Center}

\date{}

\maketitle

\noindent
\textsf{{\bf Abstract}: \input{abstract}}%

\vspace{0.3 cm}

\input{1_Introduction}

\input{3_Data_and_matched_samples}

\input{4_Notation}

\input{5_Inference}
\input{6_Simulation}
\input{7_Case_study}
\input{8_Discussion}

\bibliographystyle{apalike}
\bibliography{paper-ref}
\end{document}

%% file: abstract.tex
Regionalization of intensive care for premature babies refers to a triage system of mothers with high-risk pregnancies to hospitals of varied capabilities based on risks faced by infants. Due to the limited capacity of high-level hospitals, which are equipped with advanced expertise to provide critical care, understanding the effect of delivering premature babies at such hospitals on infant mortality for different subgroups of high-risk mothers could facilitate the design of an efficient perinatal regionalization system. Towards answering this question, Baiocchi et al. (2010) proposed to \emph{strengthen} an excess-travel-time-based, continuous instrumental variable (IV) in an IV-based, matched-pair design by switching focus to a smaller cohort amenable to being paired with a larger separation in the IV dose. Three elements changed with the strengthened IV: the study cohort, compliance rate and latent complier subgroup. Here, we introduce a non-bipartite, template matching algorithm that embeds data into a target, pair-randomized encouragement trial which maintains fidelity to the original study cohort while strengthening the IV. We then study randomization-based and IV-dependent, biased-randomization-based inference of partial identification bounds for the sample average treatment effect (SATE) in an IV-based matched pair design, which deviates from the usual effect ratio estimand in that the SATE is agnostic to the IV and who is matched to whom, although a strengthened IV design could narrow the partial identification bounds. Based on our proposed strengthened-IV design, we found that delivering at a high-level NICU reduced preterm babies' mortality rate compared to a low-level NICU for $81,766 \times 2 = 163,532$ mothers and their preterm babies and the effect appeared to be minimal among non-black, low-risk mothers.  

%% file: 1_Introduction.tex
\section{Introduction}
\label{sec: introduction}

\subsection{Does delivery at a high-level NICU reduce infant mortality?}
\label{subsec: intro clinical questions}
Regionalization of intensive care for premature babies refers to a triage system of mothers with high-risk pregnancies to hospitals of varied capabilities based on risks faced by infants. In the analyses of premature births in Pennsylvania,  \citet{baiocchi2010building} and \citet{lorch2012differential} found that delivering premature babies at a high-level neonatal intensive care unit (NICU), broadly defined as a medical unit equipped with advanced technical expertise and capacity to provide critical care for newborns and premature infants, was associated with lower infant mortality rate for a latent complier subgroup. Yet, due to the limited capacity of high-level NICUs, it is not practical to triage all mothers with high-risk pregnancies to high-level NICUs. Some previous works have further explored the treatment effect heterogeneity of delivering at a high-level versus low-level NICU. For instance, \citet{yang2014estimation} found that delivering at a high-level NICU significantly reduced deaths for babies of small gestational age but had a negligible effect for almost mature babies. More recently, \citet{chen2023estimating} estimated individual treatment rules subject to different levels of resource constraints and found that mothers' age and race/ethnicity may also be potential effect modifiers. These findings appear to align well with the clinical literature. According to a $2015$ National Vital Statistics Reports \citep{mathews2015infant}, black infants in the US had a $2.2$-fold greater mortality rate than white infants. On the other hand, it has long been established that the risk of miscarriage increases with age; in particular, \cite{hansen1986older} found that pregnant women aged $35$ or older experience an increased risk of intrauterine fetal death, pregnancy-induced hypertension, gestational diabetes, and delivery by cesarean. Motivated by these findings, we utilize observational data from the Commonwealth of Pennsylvania to investigate whether delivering preterm babies at a high-level NICU reduces infant mortality for all mothers in Pennsylvania and certain high-risk subgroups, such as black mothers and mother who had reached or were past the advanced maternal age of $35$.



\subsection{Excess travel time as an instrumental variable}
\label{subsec: intro IV PS CATE}
To estimate the treatment effect from retrospective observational data, a na\"ive study design would compare mothers who received care at a high-level NICU with those at a low-level NICU. However, as pointed out in numerous previous works \citep{baiocchi2010building, lorch2012differential,yang2014estimation,pu2021estimating,chen2023estimating}, treatment effect estimates derived from such a na\"ive comparison would be biased because receipt of neonatal intensive care at a high-level NICU as opposed to a low-level one could be easily confounded by unmeasured variables not captured in administrative databases. 

When faced with unmeasured confounding bias, researchers often resort to quasi-experimental devices (see, e.g., \citealp{cook2002experimental,rosenbaum2010design}). Instrumental variable (IV) methods are among the most popular \citep{haavelmo1943statistical,angrist1995identification,angrist1996identification}. Roughly speaking, a valid IV is a variable independent of  unmeasured treatment-outcome confounders, possibly within strata defined by observed covariates, and whose sole effect on the outcome is through its effect on the treatment (see Section \ref{sec: notation} for a formal definition in the context of a continuous IV). In a seminal paper, \citet{angrist1996identification} showed that under an additional ``monotonicity" or ``no defiers" assumption, a valid IV could be used to nonparametrically identify the treatment effect among a well-defined, albeit latent, subgroup referred to as ``compliers." 

In their original study design, \citet{baiocchi2010building} used excess travel time, or the difference in travel times (in minutes) from a mother's zip code to the nearest high-level and low-level NICUs, as an IV for delivering at a high-level versus low-level NICU. According to this definition, a small excess travel time indicates proximity to a high-level NICU relative to a low-level NICU and corresponds to a strong encouragement to deliver at a high-level NICU. Similar proximity-based IVs have been extensively used in health services research \citep{newhouse1998econometrics,baiocchi2014instrumental}. Using an IV-based method, \citet{baiocchi2010building} concluded that for every $100$ additional mothers encouraged by the IV to deliver at a high-level NICU, $0.9$ additional infant deaths could be avoided.

\subsection{Compiler average treatment effect; criticism; strengthening a continuous IV}
\label{subsec: intro complier effect and criticism}
The extensive use of instrumental variables in  empirical research is not without criticism. For instance, \citet{deaton2009instruments} argued that IV estimation methods are analogous to letting the light fall where it may when looking for an object and then proclaiming that whatever the light illuminates is what has been sought after all along. \citet{heckman2010comparing} expressed similar sentiments. The crux of their arguments is concerning the tension between the internal and external validities of IV estimates: it is, after all, not clear if the proclaimed treatment effect among compliers could be readily generalized to a target subgroup of interest. The inconvenient fact that the complier subgroup is not identified from the population and could possibly be altered by different incentive structures only makes the generalizability of IV estimates more challenging \citep{joffe2011principal}.

Responding to these criticisms, \citet{imbens2010better} argued that the internal validity of an IV estimate is often superior to that of other estimands for observational data, and that ``a credible estimate of the average effect for a subpopulation is preferred to an estimate of the average for the overall population with little credibility" \citep{imbens2010better}. As discussed by \citet{imbens2010better}, two approaches could complement an IV estimate. First, although a valid IV cannot point identify the average treatment effect (ATE), it can partially identify it; see \citet{swanson2018partial} for a recent review and references therein. Alternatively, some researchers propose additional identification assumptions that allow the complier average treatment effect to be generalized to the entire population. Some examples include the principal ignorability assumption \citep{jo2009use,angrist2010extrapolate,ding2017principal} and the homogeneity/no-interaction assumption \citep{hernan2006instruments,wang2018bounded}, among others.

In their original analysis, \citet{baiocchi2010building} used a study design technique called non-bipartite matching \citep{lu2001matching,lu2011optimal,rigdon2018near, zhang2023statistical} to construct matched pairs that comprise comparable mothers with different excess travel times. Their key innovation is to consider two study designs: one that utilized all data and created $99,174$ pairs (\textsf{Design I}) and the other that ``strengthened" the IV by only pairing two similar mothers whose IVs were far apart, yielding $49,587$ pairs (\textsf{Design II}). 

Three things changed between \textsf{Design I} and \textsf{Design II}. First, the population under investigation changed; for instance, the proportion of white mothers increased from $70\%$ in \textsf{Design I} to $85\%$ in \textsf{Design II} \citep[Table 1]{baiocchi2010building}. Second, the compliance rate changed. \citeauthor{baiocchi2010building}'s \citeyearpar{baiocchi2010building} key insight is that, by forcing the continuous IV to be further apart within each matched pair, the compliance rate induced by this strengthened IV increased, improving statistical inference of the effect ratio estimand, an analogue of the Wald estimate in a matched-pair design. Third, the latent complier subgroup changed. The original IV in \textsf{Design I} and its strengthened version in \textsf{Design II} targeted a sample average treatment effect (SATE) over different complier subgroups. \citet[Section 5]{baiocchi2010building} argued that the effect ratio estimate obtained from the strengthened design was more ``typical" in the sense that a typical mother was more likely to respond to a strong incentive in \textsf{Design II} as opposed to a much weaker incentive in \textsf{Design I}. 

\subsection{Manipulating a continuous IV with two purposes; caveats}
\label{subsec: manipulate IV}
In his perspective piece, \citet{deaton2009instruments} argued that standard statistical practice would first define a parameter of interest and then construct an estimator targeting the parameter. In the causal inference literature, a causal estimand of interest is often defined first, followed by identification assumptions and estimation procedures operated under these identification assumptions. On the contrary, IV-based procedures implicitly let the chosen IV, which may have little to do with the scientific question at hand, determine the causal estimand.

In this article, we propose a study design framework that builds upon \citeauthor{baiocchi2010building}'s \citeyearpar{baiocchi2010building} key insight and show that an ideal paradigm---stating a well-defined causal estimand for an identifiable target subgroup and \emph{irrespective of} the IV, followed by identification and estimation using an IV---could be achieved by manipulating a continuous IV in a design-based framework \citep{rosenbaum2002covariance, imbens2005robust}. Our proposed study design starts by building matched pairs that satisfy the following three criteria: (i) the covariates of matched mothers are deemed well balanced according to formal diagnostic tests; (ii) the covariate distribution of the matched sample closely resembles that of a target study population; and (iii) the within-pair difference in the IV is maximized to the extent that the data can afford, with the goal of maximizing the IV-induced compliance rate. We then propose a randomization inference (RI) approach to infer partial identification bounds of the sample average treatment effect and extend the approach to accommodate an IV-dose-dependent, biased randomization scheme. Our framework deviates from  previous works that conduct RI-based inference for the structural parameter in a constant, proportional treatment effect model \citep{imbens2005robust, small2008war} or the effect ratio estimand \citep{baiocchi2010building,zhang2022bridging}. By strengthening the IV and maximizing the compliance rate, a careful design could help derive more informative bounds for the desired effect. In \citeauthor{deaton2009instruments}'s \citeyearpar{deaton2009instruments} analogy, researchers clearly state what is desired to be sought after and then adjust the brightness of the light (i.e., strengthen the continuous IV) to illuminate the object. 

A valid, continuous IV is at the heart of our proposed design. A binary IV cannot be strengthened; it appears that researchers have to accept the IV as it is and its associated complier subgroup. Fortunately, a lot of commonly used IVs, such as variables related to geography and nature, are often continuous. Alternatively, one may combine multiple binary IVs to construct a many-category IV that is amenable to being strengthened. An important caveat is regarding the validity of the IV: if the putative IV is not valid, then strengthening it could increase the bias if the unmeasured IV-outcome confounder is correlated with the continuous IV \citep{heng2023instrumental}.

%% file: 3_Data_and_matched_samples.tex
\section{Data and study design}
\label{sec: data and design}
\subsection{NICU data; target trial; design considerations}
\label{subsec: data and design and alg}
We considered data on $181,762$ preterm babies in the Commonwealth of Pennsylvania between 1995 and 2004. As stipulated by the American Academy of Pediatrics, there are six levels of NICUs of increasing technical expertise and capability: 1, 2, 3A, 3B, 3C and 3D, as well as regional centers 4 \citep{baiocchi2010building}. We followed \citet{baiocchi2010building} and defined a hospital to be high level if it delivered at least 50 preterm babies per year on average and its NICU was of level 3A-3D or 4; we defined a hospital as low level if it delivered fewer than 50 preterm babies annually or its NICU was below level 3A. 

Our goal is to embed the retrospective observational data into a hypothetical randomized encouragement trial in such a way that the instrument is strengthened to the extent that the data can afford and the final matched sample closely resembles the template. Three elements are key to our proposed study design. First, mothers with similar observed covariates but different IV-defined exposures, e.g., excess travel times, need be paired together. One approach would be to dichotomize the continuous exposure and adopt a conventional matching algorithm designed for a binary exposure \citep{rosenbaum2020modern}; however, this practice leads to a potential violation of the stable unit treatment value assumption (SUTVA) \citep{rubin1980randomization,rubin1986statistics}.  As an alternative, non-bipartite matching creates matched pairs without dichotomizing the continuous exposure \citep{lu2001matching,lu2011optimal,baiocchi2010building,zhang2023statistical}. Second, when the data can afford, it is useful to maximize the within-matched-pair difference in the IV dose so that the compliance rate can be maximized. Third, we would like the final matched sample to closely resemble the template, i.e., a random sample from a target subgroup, to enhance the generalizability of the clinical conclusions \citep{bennett2020building,zhang2023efficient}. 

To accommodate these three design considerations, we modified the optimal non-bipartite matching algorithm in \citet{baiocchi2010building}. Our key idea is to add \textit{e} ``phantom units," or ``sinks," that represent units in the template and design a discrepancy matrix in such a way that study participants who do not resemble the template are matched to the sinks and thus eliminated from the final matched sample. Suppose there are $n$ mothers to be matched and $e < n$ sinks representing a target template of mothers. Define the following $(n+e) \times (n+e)$ distance matrix:
$$\mathcal{M} = \left( \begin{array} {cccc|ccc}
\infty & \delta_{1,2} & \cdots & \delta_{1,n} & \Delta_{1, n+1} & \cdots & \Delta_{1, n+e} \\
\delta_{2,1} & \infty & \cdots & \delta_{2,n} & \Delta_{2, n+1} & \cdots & \Delta_{2, n+e} \\
\vdots & & \ddots & \vdots & \vdots & \ddots & \vdots \\
\delta_{n,1} & \delta_{n,2} & \cdots & \infty & \Delta_{n, n+1} & \cdots & \Delta_{n, n+e} \\
\hline
\Delta_{n+1, 1} & \Delta_{n+1, 2} & \cdots & \Delta_{n+1, n} & \infty & \cdots & \infty \\
\vdots & & \ddots & \vdots & \infty & \ddots & \infty \\
\Delta_{n+e, 1} & \Delta_{n+e, 2} & \cdots & \Delta_{n+e, n} & \infty & \cdots & \infty \\
\end{array}\right),$$
where $\mathcal{M}_{i,j}$ denotes the $(i,j)$-th entry of $\mathcal{M}.$ Here, $\mathcal{M}_{i,j}$ represents a distance between two mothers to be matched for $i,j \in [n], $ between two mothers in the template for $i,j \in \{n+1, n+2, \dots, n+e\},$ and between a mother to be matched and a mother in the template for the remaining entries. 

The discrepancy matrix was built based on the following steps. First, for $i, j \in [n]$, we let $\delta_{i,j}$ be a measure of the distance between the observed covariates $x_i$ and $x_j$ of mothers $i$ and $j,$ respectively. Then $\delta_{i,j}$ measures homogeneity between two mothers and can take different specifications. A most popular choice is the (rank-based) Mahalanobis distance within the estimated propensity score caliper \citep{rosenbaum1985constructing}. To strengthen the continuous IV, we additionally incorporated the dose caliper to force excess travel times within each matched pair to be far apart, thus strengthening the IV and improving the IV-induced compliance rate \citep[Section 5]{zhang2023statistical}. Second, for $i \in [n]$ and $j \in \{n+1, n+2, \dots, n+e\},$ let $\Delta_{i,j}$ be a measure of the distance between the observed covariates of mother $i$ to be matched and mother $j$ in the template. Intuitively, if mother $i$ is dissimilar to mother $j$, we would want to pair them together, in which case $\Delta_{i,j}$ would need to be small, in order to eliminate mother $i$ from the final matched sample. To this end, we let $\Delta_{i,j}$ equal an arbitrary large number minus the (rank-based) Mahalanobis distance between $x_i$ and $x_j$. We also incorporated a ``reverse" caliper on the estimated probability of being selected into the template given observed covariates, an analogue of the propensity score \citep{cole2010generalizing,stuart2011use}. Lastly, we set $\mathcal{M}_{i,j} = \infty$ for $i, j \in \{n+1, n+2, \dots, n+e\}$ to prevent two template mothers from being matched together.

After a discrepancy matrix $\mathcal{M}$ is constructed, an optimal non-bipartite matching algorithm takes it as an input and yields $(n+e)/2$ matched pairs that minimize the sum of the discrepancies. The final matched sample comprises $(n-e)/2$ pairs of observational units and the remaining $e$ pairs consisting of an observational unit and a phantom unit. The algorithm runs in polynomial time and is available in the \textsf{R} package \textsf{nbpmatching} \citep{derigs1988solving,lu2011optimal}.

\subsection{Two strengthened matched comparisons}
\label{data: matched sample}
We constructed and compared two matched samples, one constructed using a usual technique as in the original \citeauthor{baiocchi2010building}'s \citeyearpar{baiocchi2010building} paper ($\textsf{M0}$) and the other based on the algorithm proposed in Section \ref{subsec: data and design and alg} ($\textsf{M1}$). For both designs, we matched mothers on birth weight, parity, insurance type (fee for service or others), below poverty, mother's age, mother's education, and a single birth indicator, and matched exactly on the year the data was collected. For subgroup analyses based on $\textsf{M1}$, we further matched on a categorical, high-risk variable defined by a combination of a mom's race, age, and gestational age, motivated by clinical literature discussed in Section \ref{subsec: intro clinical questions}: the high-risk category equals $0$ if a mother is black, $1$ if a mother is at least
$35$ years old and her gestational age does not exceed $36$, $2$ if a mother is younger than $35$ and her gestational age does not exceed $36$, and $3$ if a mother's gestational age exceeds $36$. 

The first matched sample \textsf{M0} retained and matched all $181,722$ mothers into pairs using a conventional non-bipartite matching algorithm \citep{lu2001matching,lu2011optimal}. For the second matched sample \textsf{M1}, we first created a template subgroup of mothers by randomly sampling without replacement $10$ percent of mothers from the entire dataset spanning $1995$ to $2004$. We then implemented our statistical matching algorithm, which includes several design devices such as a generalizability caliper and an IV dose caliper. Design \textsf{M1} yielded $81,766$ matched pairs of mothers whose within-matched-pair rank-based Mahalanobis distance was minimized subject to the constraints enforced by the above generalizability and IV dose calipers. See Supplemental Materials C for details.

Table \ref{tb: covariate balance} contrasts the average excess travel times and covariate balance between mothers encouraged to deliver at high-level and low-level NICUs in the matched samples \textsf{M0} and \textsf{M1}. In both matches, the absolute SMD of almost every covariate between matched pairs of mothers does not exceed $0.1$, or one-tenth of one standard deviation \citep{silber2001matching}. We will formally test the randomization and relaxed randomization assumption before conducting inference in Section \ref{sec: case study}. 


\begin{table}[!ht]
\centering
\caption{Degree of encouragement and covariate balance between mothers encouraged to deliver at high-level (near) versus low-level (far) NICUs in two matched comparisons using the $1995-2004$ data. Matches \textsf{M0} and \textsf{M1} are created without and with a dose caliper, respectively. For each variable, we report the mean and the absolute standardized difference. GA means gestational age.}
\label{tb: covariate balance}
\scalebox{0.85}{
\begin{tabular}{lcccccc} \toprule
& \multicolumn{3}{c}{Match 0 ($\textsf{M0}$)} & \multicolumn{3}{c}{Match 1 ($\textsf{M1}$)}
\\ 
& \multicolumn{3}{c} {$90,861$ matched pairs} & \multicolumn{3}{c}{$81,766$ matched pairs} \\
\cmidrule(lr){2-4} \cmidrule(lr){5-7} 
& Near   & Far & Absolute & Near & Far  & Absolute \\
& Mean  & Mean & SMD    & Mean &  Mean & SMD \\
\midrule
\textbf{Instrumental variable} \\
~~Excess travel time (min) & 6.07 & 20.97 & 0.93 & 2.07 & 27.10 & 1.88 \\ 
\textbf{Covariates}\\
~~Birthweight (g) & 2583 & 2584 & 0.00 & 2597 & 2597 & 0.00 \\ 
~~Gestational age (weeks) & 35.12 & 35.13 & 0.00 & 35.21 & 35.19 & 0.01 \\ 
~~Gestational diabetes (0/1) & 0.05 & 0.05 & 0.01 & 0.05 & 0.05 & 0.01 \\ 
~~Single birth (0/1) & 0.83 & 0.83 & 0.00 & 0.85 & 0.84 & 0.02 \\ 
~~Parity & 2.11 & 2.12 & 0.01 & 2.06 & 2.11 & 0.03 \\ 
~~Mother's age (years) & 28.05 & 28.04 & 0.00 & 28.03 & 27.80 & 0.04 \\ 
~~Mother's education (scale) & 3.69 & 3.68 & 0.00 & 3.73 & 3.63 & 0.08 \\ 
~~Black (0/1) & 0.16 & 0.16 & 0.00 & 0.16 & 0.16 & 0.00 \\ 
~~Below poverty (proportion) & 0.13 & 0.13 & 0.01 & 0.12 & 0.12 & 0.06 \\ 
~~Fee for service (0/1) & 0.21 & 0.21 & 0.00 & 0.20 & 0.21 & 0.05 \\ 
~~High risk category & \\ 
~~~~Black & 0.16 & 0.16 & 0.00 & 0.16 & 0.16 & 0.00 \\
~~~~Non-Black, age $\geq 35$, GA $\leq 36$ & 0.08 & 0.08 & 0.00 & 0.08 & 0.08 & 0.00 \\
~~~~Non-Black, age $< 35$, GA $\leq 36$ & 0.40 & 0.39 & 0.00 & 0.40 & 0.40 & 0.00 \\
~~~~Non-Black, GA $> 36$ & 0.36 & 0.36 & 0.00 & 0.36 & 0.36 & 0.00 \\
\bottomrule
\end{tabular}
}
\end{table}

A key difference between $\textsf{M0}$ and $\textsf{M1}$ is the level of separation in the IV enforced by the study design. The average differences (larger minus smaller within each matched pair) in excess travel times between mothers encouraged to deliver at a low-level versus a high-level NICU for \textsf{M0} and \textsf{M1} are $20.97 - 6.07 = 14.90$ and $27.10 - 2.07 = 25.03$ minutes, respectively. This is in line with our study design: a large IV dose caliper enforces the excess travel times for mothers in each matched pair to be further apart, thereby strengthening the IV and improving the IV-induced compliance rate from $65.4\% - 45.6\% = 19.8\%$ in \textsf{M0} to $71.9\% - 35.5\% = 36.4\%$ in \textsf{M1}. According to the analytical results in \citet{heng2023instrumental}, in order for $\textsf{M0}$ to attain the same power as $\textsf{M1}$ when testing Fisher's sharp null hypothesis, $\textsf{M0}$ needs to have a sample size that is $(36.4\%/19.8\%)^2 = 3.38$ times that of $\textsf{M1}$. Design \textsf{M0}'s sample size is only $1.1$ times that of $\textsf{M1}$; therefore, from the perspective of testing Fisher's sharp null hypothesis, the strengthened design $\textsf{M1}$ is more advantageous. 

\begin{figure}[!ht]
\caption{Boxplots of the rank-based Mahalanobis distances of two types of matched pairs in \textsf{M1}: (i) two observational units (Obs-Obs) or (ii) an observational unit and a template unit (Obs-Template).}
\label{fig: matched distance}
\centering
\includegraphics[width=0.8\textwidth]{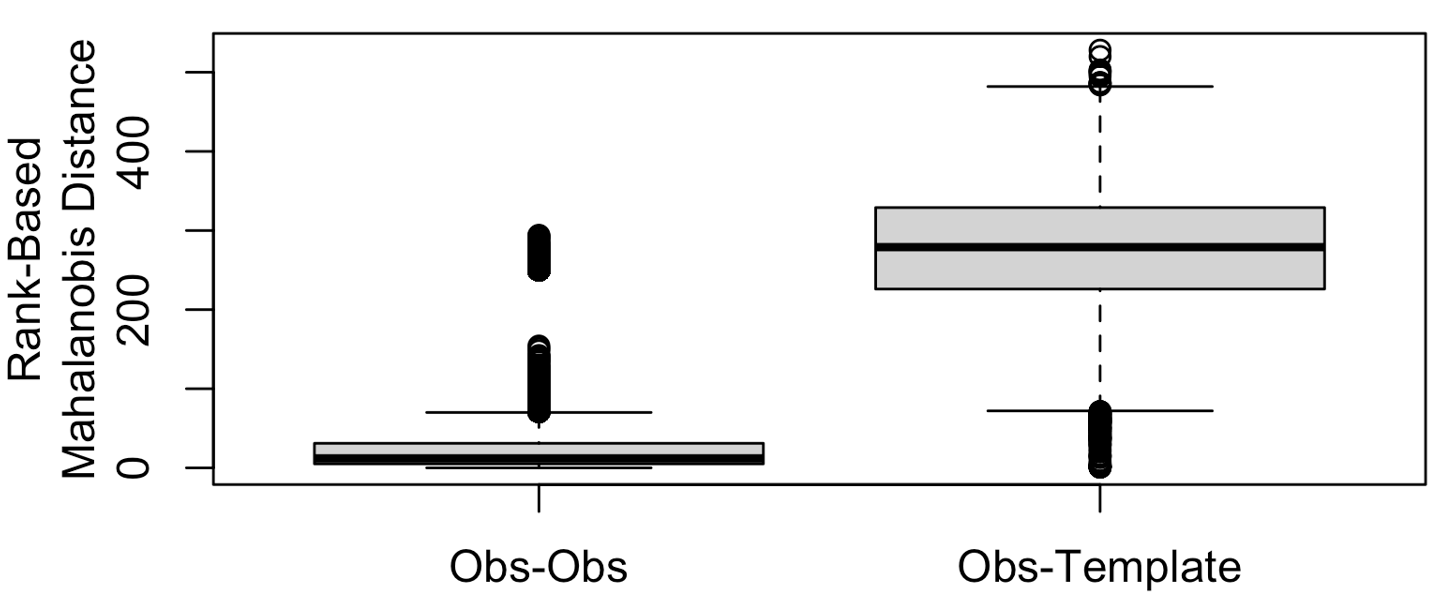}
\end{figure}

Furthermore, by using the template matching technique and enforcing the generalizability caliper, the matched sample \textsf{M1} also exhibits good generalizability to the template. For example, \textsf{M1} largely retains the original composition of black mothers, who comprise $16\%$  of all the mothers in the observational data, and of mothers with fee-for-service health insurance. This contrasts the original strengthened IV design in \cite{baiocchi2010building} which, in transitioning from the unstrengthened to strengthened IV matching design, shifted the study cohort from $15\%$ to $5\%$ of black mothers and from $21\%$ to $25\%$ of mothers with fee-for-service health insurance. Figure \ref{fig: matched distance}, which illustrates the distribution of rank-based Mahalanobis distances between matched pairs retained in or eliminated from \textsf{M1}, further demonstrates the effectiveness of our matching algorithm in maintaining good generalizability of the final matched sample to the entire study cohort. For the matched pairs that each consist of two mothers from the observational database and thus comprise \textsf{M1}, the median rank-based Mahalanobis distance is $12$ (interquartile range: [5, 31]). For the matched pairs that are eliminated from \textsf{M1}, the median rank-based Mahalanobis distance is $279$ (interquartile range: [226, 329]). This is in line with our expectation, as mothers who resembled the template the least in terms of the covariates had large rank-based Mahalanobis distances when matched to the template and were thus eliminated from the final matched sample \textsf{M1}.


%% file: 4_Notation.tex
\section{Notation: potential outcomes and estimands}
\label{sec: notation}
\subsection{Potential outcomes}
\label{subsec: PO}
Our proposed design consists of $I$ matched pairs of 
two study participants. The $j$-th participant in the $i$-th pair is indexed by $ij$. Each participant $ij$ is associated with a continuous IV  $\widetilde{Z}_{ij} = \widetilde{Z}_{ij}^{\text{obs}}$ e.g., excess travel time in the NICU study. The design embeds the observational data into a hypothetical randomized encouragement experiment as follows. Within each matched pair $i$, fix the two doses of the continuous IV at $\widetilde{Z}_{i1}^{\text{obs}}$ and $\widetilde{Z}_{i2}^{\text{obs}}$ and flip a fair coin. If the coin lands heads, assign the IV dose $\widetilde{Z}_{i1}^{\text{obs}}$ to $j = 1$ and $\widetilde{Z}_{i2}^{\text{obs}}$ to $j = 2$; if the coin lands tails, assign the IV dose $\widetilde{Z}_{i1}^{\text{obs}}$ to $j = 2$ and $\widetilde{Z}_{i2}^{\text{obs}}$ to $j = 1$. Let $\widetilde{\mathbf{Z}}=(\widetilde{Z}_{11}, \dots, \widetilde{Z}_{I2})$ denote the vector of $I \times 2 = 2I$ IV dose assignments. In a randomized encouragement experiment, it is the encouragement (i.e., the continuous IV $\widetilde{Z}$), not the treatment itself, that is randomized either by nature or an experimenter.

Let $D_{ij}(\widetilde{\mathbf{Z}})$ denote the indicator of whether participant $ij$ receives the binary treatment under the IV dose vector $\widetilde{\mathbf{Z}}$. We make the SUTVA assumption \citep{rubin1980randomization,rubin1986statistics} so that $D_{ij}(\widetilde{\mathbf{Z}})$ depends on $\widetilde{\mathbf{Z}}$ only via $\widetilde{Z}_{ij}$ such that $D_{ij}(\widetilde{\mathbf{Z}})$ can be simplified to $D_{ij}(\widetilde{Z}_{ij})$. We adopt the notation in \citet{zhang2022bridging} and \citet{heng2023instrumental} and define the following potential outcomes:
\begin{equation}
\label{eq: definition of dt and dc}
    \begin{split}
         &d_{Cij}\overset{\Delta}{=} D_{ij}( \widetilde{Z}_{ij}=\widetilde{Z}_{i1}^\text{obs}\wedge \widetilde{Z}_{i2}^\text{obs})~ \quad \text{and} \quad~  d_{Tij}\overset{\Delta}{=} D_{ij}( \widetilde{Z}_{ij}=\widetilde{Z}_{i1}^\text{obs}\vee \widetilde{Z}_{i2}^\text{obs}),
    \end{split}
\end{equation}
where $\widetilde{Z}_{i1}^\text{obs}\wedge \widetilde{Z}_{i2}^\text{obs}$ denotes the minimum of two IV doses and $\widetilde{Z}_{i1}^\text{obs}\vee \widetilde{Z}_{i2}^\text{obs}$ the maximum of two IV doses in the $i$-th matched pair. Importantly, participant $ij$'s compliance status is determined by two IV doses in matched pair $i$ and the compliance status is fixed once the study design is fixed. Participant $ij$ is said to be a complier with respect to the IV dose pair $(\widetilde{Z}_{i1}^{\text{obs}}\wedge \widetilde{Z}_{i2}^{\text{obs}}, \widetilde{Z}_{i1}^{\text{obs}}\vee \widetilde{Z}_{i2}^{\text{obs}})$ if $(d_{Tij}, d_{Cij}) = (1, 0)$, an always-taker if $(d_{Tij}, d_{Cij}) = (1, 1)$, a never-taker if $(d_{Tij}, d_{Cij}) = (0, 0)$, and a defier if $(d_{Tij}, d_{Cij}) = (0, 1)$, where $d_{Tij}$ and $d_{Cij}$ are defined as in \eqref{eq: definition of dt and dc}. We make the so-called monotonicity assumption \citep{angrist1996identification} and assume there are no defiers. This is often a mild assumption as it merely states that if unit $ij$ accepts the treatment under dose $\widetilde{Z}$, then it will also accept the treatment under any dose $\widetilde{Z}^\prime$ such that $\widetilde{Z}^\prime > \widetilde{Z}$. Note that even under the monotonicity assumption, researchers can still only observe one of the two potential treatments received, so statistical inference is needed. Under the monotonicity assumption, the compliance rate in the matched design with respect to a continuous IV is then defined as 
\begin{equation}
    \label{eq: compliance}
    \iota_C = \frac{1}{2I}\sum_{i = 1}^I \sum_{j = 1}^2 (d_{Tij} - d_{Cij}),
\end{equation}
and a strengthened-IV design improves the compliance rate by maximizing the IV dose difference within each matched pair \citep{baiocchi2010building}.


Let $R_{ij}(\widetilde{\mathbf{Z}}, \mathbf{D})$ denote the outcome of $ij$ under $\widetilde{\mathbf{Z}}$ and $\mathbf{D} = (D_{11}, \dots, D_{I2})$. We assume SUTVA again so that $R_{ij}(\widetilde{\mathbf{Z}}, \mathbf{D})$ depends on $\widetilde{\mathbf{Z}}$ and $\mathbf{D}$ only via $\widetilde{Z}_{ij}$ and $D_{ij}$. Lastly, we make the exclusion restriction assumption so that the effect of $\widetilde{Z}_{ij}$ on $R_{ij}$ is only via $D_{ij}$; in other words, $R_{ij}(\widetilde{Z}_{ij}, D_{ij})$ can be written as $R_{ij}(D_{ij}(\widetilde{Z}_{ij}))$ or $R_{ij}(\widetilde{Z}_{ij})$ as a shorthand. Define the following potential outcomes:
\begin{equation}
\label{eq: definition of rt and cc}
    \begin{split}
         &r_{Cij}\overset{\Delta}{=} R_{ij}( \widetilde{Z}_{ij}=\widetilde{Z}_{i1}^\text{obs}\wedge \widetilde{Z}_{i2}^\text{obs})~ \quad \text{and} \quad~  r_{Tij}\overset{\Delta}{=} R_{ij}( \widetilde{Z}_{ij}=\widetilde{Z}_{i1}^\text{obs}\vee\widetilde{Z}_{i2}^\text{obs}),
    \end{split}
\end{equation}
in parallel to $(d_{Tij}, d_{Cij})$ in \eqref{eq: definition of dt and dc}. Analogous to $(d_{Tij}, d_{Cij})$, $(r_{Tij}, r_{Cij})$ are defined with respect to the continuous IV dose pair $(\widetilde{Z}_{i1}^\text{obs}\wedge \widetilde{Z}_{i2}^\text{obs}, \widetilde{Z}_{i1}^\text{obs}\vee \widetilde{Z}_{i2}^\text{obs})$. In the NICU study, $r_{Cij}$ describes the potential infant mortality status had the mother lived $\widetilde{Z}_{ij}=\widetilde{Z}_{i1}^\text{obs}\wedge \widetilde{Z}_{i2}^\text{obs}$ (as opposed to $\widetilde{Z}_{ij}=\widetilde{Z}_{i1}^\text{obs}\vee \widetilde{Z}_{i2}^\text{obs}$) minutes farther away from a high-level NICU compared to a low-level NICU and hence encouraged to attend a high-level NICU, while $r_{Tij}$ describes the potential infant mortality status had the mother lived $\widetilde{Z}_{ij}=\widetilde{Z}_{i1}^\text{obs}\vee \widetilde{Z}_{i2}^\text{obs}$ (as opposed to $\widetilde{Z}_{ij}=\widetilde{Z}_{i1}^\text{obs}\wedge \widetilde{Z}_{i2}^\text{obs}$) minutes farther away from a high-level NICU compared to a low-level NICU and hence encouraged to attend a low-level NICU. Importantly, in either case, no mother is forced to attend any hospital. Potential outcomes $(r_{Tij}, r_{Cij})$ are fixed once the design and hence the two IV doses $(\widetilde{Z}_{i1}^{\text{obs}}\wedge \widetilde{Z}_{i2}^{\text{obs}}, \widetilde{Z}_{i1}^{\text{obs}}\vee \widetilde{Z}_{i2}^{\text{obs}})$ within each matched pair are fixed. 

\subsection{Effect ratio estimand; sample average treatment effect}
Equipped with these potential outcomes, we now turn to causal estimands. \citet{baiocchi2010building} consider the following effect ratio estimand:
\begin{equation*}
    \lambda = \frac{\sum_{i = 1}^I \sum_{j = 1}^2 (r_{Tij} - r_{Cij})}{\sum_{i = 1}^I \sum_{j = 1}^2 (d_{Tij} - d_{Cij})},
\end{equation*}
which is the ratio of the sample average treatment effect of the IV (possibly after strengthening) on the outcome and that of the IV on the treatment. The effect ratio estimand $\lambda$ is always well defined provided that the denominator $\sum_{i = 1}^j \sum_{j = 1}^2 (d_{Tij} - d_{Cij}) \neq 0$, and it coincides with the Wald estimator under an additional monotonicity assumption. In the NICU study, the effect ratio can be interpreted as follows: for every hundred mothers discouraged by the IV from delivering at a high-level NICU, there were $100 \times \lambda$ additional infant deaths \citep{baiocchi2010building}. As researchers switch to a different IV-based matched design, such as one that further separates the two IV doses within each matched pair, the effect ratio estimand will in general be different. In other words, \emph{the effect ratio estimand depends on who is matched to whom, and the design dictates the estimand.}

Lastly, we define the following potential outcomes:
\begin{equation*}
    \begin{split}
         &r_{d = 0, ij}\overset{\Delta}{=} R_{ij}(D_{ij} = 0)~ \quad \text{and} \quad~  r_{d = 1, ij}\overset{\Delta}{=} R_{ij}(D_{ij} = 1),
    \end{split}
\end{equation*}
 which correspond to potential binary infant mortality outcomes had mother $ij$ attended a low-level NICU ($D_{ij} = 1$) or a high-level NICU ($D_{ij} = 0$). Compared to $r_{Tij} - r_{Cij}$ that describes the unit-level intention-to-treat effect (with respect to the designed IV), $r_{d = 1, ij} - r_{d = 0, ij}$ describes the unit-level treatment effect and arguably is the ultimate estimand of interest in the NICU study. Below, we will focus on the sample average treatment effect of $D$ on $R$:
\begin{equation*}
    \kappa = \frac{1}{2I} \sum_{i = 1}^I\sum_{j = 1}^2 (r_{d = 1, ij} - r_{d = 0, ij}).
\end{equation*}
Compared to the effect ratio estimand $\lambda$, the estimand $\kappa$ depends on the study cohort but not on the strength of the IV or who is matched to whom in the design phase.


\subsection{Partial identification bounds}
\label{subsec: inference}
The estimand $\kappa$ cannot in general be point identified, as data provides no treatment effect information for non-compliant participants. Partial identification bounds provide an assumption-lean alternative. Define the following index sets: $I_{\text{COM}} = \{(i,j): (d_{Tij}, d_{Cij}) = (1, 0)\}$, $I_{\text{AT}} = \{(i,j): (d_{Tij}, d_{Cij}) = (1, 1)\}$, and $ I_{\text{NT}} = \{(i,j): (d_{Tij}, d_{Cij}) = (0, 0)\}$. Note that 
\begin{equation*}
\begin{split}
    \kappa = \frac{1}{N} \sum_{ij} (r_{d = 1, ij} - r_{d = 0, ij}) = &\frac{1}{N}\left\{\sum_{(i,j) \in I_{\text{COM}}} (r_{d = 1, ij} - r_{d = 0, ij}) + \sum_{(i,j) \in I_{\text{AT}} \cup I_{\text{NT}} } (r_{d = 1, ij} - r_{d = 0, ij})\right\},
\end{split}
\end{equation*}
where the first term can be further simplified to
\begin{equation*}
    \sum_{(i,j) \in I_{\text{COM}}} (r_{d = 1, ij} - r_{d = 0, ij}) = \sum_{(i,j) \in I_{\text{COM}}} (r_{Tij} - r_{Cij}) = \sum_{ij} (r_{Tij} - r_{Cij}).
\end{equation*}



Let $r_{d = 1, ij} - r_{d = 0, ij} \in [K_0, K_1]$ for all $ij$ and for some $K_0 \leq K_1 < \infty.$ Without loss of generality, assume $K_0 \geq 0$. The second term is then bounded as follows:
\begin{equation*}
     K_0 \times (N - |I_{\text{COM}}|) \leq \sum_{(i,j) \in I_{\text{AT}} \cup I_{\text{NT}} } (r_{d = 1, ij} - r_{d = 0, ij}) \leq K_1 \times (N - |I_{\text{COM}}|).
\end{equation*}
Because $\sum_{ij} (d_{Tij} - d_{Cij}) = \sum_{(i,j) \in I_{\text{COM}}} (d_{Tij} - d_{Cij}) = |I_{\text{COM}}|$, we then have
\begin{equation}
\label{eq: K0<<K1}
     K_0 \times \left(N - \sum_{ij} (d_{Tij} - d_{Cij})\right) \leq \sum_{(i,j) \in I_{\text{AT}} \cup I_{\text{NT}} } (r_{d = 1, ij} - r_{d = 0, ij}) \leq K_1 \times \left(N - \sum_{ij} (d_{Tij} - d_{Cij})\right).
\end{equation}
Put together, without additional assumptions, the target estimand $\kappa$ is lower bounded by
\begin{equation*}
   \text{LB} := \frac{1}{N} \sum_{ij} (r_{Tij} - r_{Cij}) - K_0 \times \frac{1}{N} \sum_{ij} (d_{Tij} - d_{Cij}) + K_0, 
\end{equation*}
and upper bounded by 
\begin{equation*}
   \text{UB} := \frac{1}{N} \sum_{ij} (r_{Tij} - r_{Cij}) - K_1 \times \frac{1}{N} \sum_{ij} (d_{Tij} - d_{Cij}) + K_1. 
\end{equation*}
\noindent Three remarks follow. First, the lower and upper bounds can be achieved when the treatment effect among all the non-compliers attain the minimum and maximum ($K_0$ and $K_1$, respectively). Second, the length of the partial identification interval, $\text{UB} - \text{LB}$, equals $(K_1 - K_0) \times (1 - \iota_C)$, where $\iota_C$ is the compliance rate defined in \eqref{eq: compliance}. Thus, by maximizing the compliance rate, a strengthened-IV design could help derive a narrower partial identification interval for $\kappa$. In an ideal situation where $\iota_C = 1$, we have $\text{LB} = \text{UB}$ and the identification of $\kappa$ is achieved by design. Third, by dividing $N - \sum_{ij} (d_{Tij} - d_{Cij})$ in equation \eqref{eq: K0<<K1}, $K_0$ and $K_1$ can also be interpreted as the lower and upper bounds of the SATE among non-compliers. 


The target parameters LB and UB involve both potential treatments received $(d_{Tij}, d_{Cij})$ and  both potential clinical outcomes $(r_{Tij}, r_{Cij})$; for each study unit $ij$, either $(d_{Tij}, r_{Tij})$ or $(d_{Cij}, r_{Cij})$ is observed, so inference is needed for the LB and UB.

%% file: 5_Inference.tex
\section{Inference for partial identification bounds}
\label{sec: RI}

\subsection{Inference under a randomization assumption}
\label{subsec: inference under RI}
One widely-adopted downstream analysis strategy for matched cohort data is randomization inference treating the matched cohort data as from a finely stratified experiment \citep{rosenbaum2002observational,rosenbaum2010design,fogarty2018mitigating,fogarty2018regression}. We will focus on developing a valid statistical test and confidence statement for the partial identification bounds under the randomization assumption in this section; methods that further take into account within-pair residual bias that persists after matching \citep{gagnon2019cpt,chen2023testingbiased} will be discussed next. 

Write $\mathcal{F}=\{(\mathbf{x}_{ij}, d_{Tij}, d_{Cij}, r_{Tij}, r_{Cij}): i=1,\dots, I, j=1,2 \}$, where $d_{Tij}$, $d_{Cij}$, $r_{Tij}$ and $r_{Cij}$ are defined as in \eqref{eq: definition of dt and dc} and \eqref{eq: definition of rt and cc}.
Write $\mathcal{Z}$ for the set containing the $2^I$ possible values $\widetilde{\mathbf{z}}$ of IV dose assignments $\widetilde{\mathbf{Z}}$, so that $\widetilde{\mathbf{z}} \in \mathcal{Z}$ if each $\widetilde{{z}}_{ij}$ equals $\widetilde{Z}^{\text{obs}}_{i1} \vee \widetilde{Z}^{\text{obs}}_{i2}$ or $\widetilde{Z}^{\text{obs}}_{i1} \wedge \widetilde{Z}^{\text{obs}}_{i2}$. 
Let $\widetilde{\mathbf{Z}}^{\text{obs}}_{\vee}=(\widetilde{Z}^{\text{obs}}_{11}\vee \widetilde{Z}^{\text{obs}}_{12}, \dots, \widetilde{Z}^{\text{obs}}_{I1}\vee \widetilde{Z}^{\text{obs}}_{I2})$ and $\widetilde{\mathbf{Z}}^{\text{obs}}_{\wedge}=(\widetilde{Z}^{\text{obs}}_{11}\wedge \widetilde{Z}^{\text{obs}}_{12}, \dots, \widetilde{Z}^{\text{obs}}_{I1}\wedge \widetilde{Z}^{\text{obs}}_{I2})$. In randomization inference with a continuous IV $\widetilde{\mathbf{Z}}$, the only probability distribution that enters statistical inference is the conditional probability $\pi_{i} = P(\widetilde{Z}_{i1} = \widetilde{Z}^{\text{obs}}_{i1} \vee \widetilde{Z}^{\text{obs}}_{i2}, \widetilde{Z}_{i2} = \widetilde{Z}^{\text{obs}}_{i1}\wedge  \widetilde{Z}^{\text{obs}}_{i2}\mid \mathcal{F}, \widetilde{\mathbf{Z}}^{\text{obs}}_{\vee}, \widetilde{\mathbf{Z}}^{\text{obs}}_{\wedge})$ that characterizes the underlying IV dose assignment mechanism in each matched pair $i$. Assuming the IV is valid and the randomization assumption holds, we have $\pi_i = 1/2$ for $i = 1, \cdots, I$. Define a shorthand $Z_{ij} = 1$ if $\widetilde{Z}_{ij}=\widetilde{Z}_{i1}^\text{obs}\vee \widetilde{Z}_{i2}^\text{obs}$, $Z_{ij} = 0$ if $\widetilde{Z}_{ij}=\widetilde{Z}_{i1}^\text{obs}\wedge \widetilde{Z}_{i2}^\text{obs}$ and $\mathbf{Z} = (Z_{11}, Z_{12}, \dots, Z_{I2})$. 
Under the matched pair design  conditional on $\mathcal{Z}$, $Z_{i1}+Z_{i2}=1$ for each $i$. 
Let $R_{ij}=Z_{ij}r_{Tij}+(1-Z_{ij})r_{Cij}$ denote the observed outcome and $D_{ij}=Z_{ij}d_{Tij}+(1-Z_{ij})d_{Cij}$ denote the observed treatment received. 
We consider the null hypothesis $H_0^L: \text{LB} = l$ and an ever-larger experiment with $I \rightarrow \infty$. Proposition \ref{prop: inference} derives an asymptotically valid test for $H_0^L$.

\begin{proposition}\label{prop: inference}
Define the test statistic
\[
T\left(l;K_0\right)
 =\frac{1}{I} \sum_{i=1}^I\left\{\sum_{j=1}^2 Z_{i j}\left(R_{i j}-K_0 D_{i j}\right)-\sum_{j=1}^2\left(1-Z_{i j}\right)\left(R_{i j}-K_0 D_{i j}\right)\right\} - (l-K_0)
\] and $
S^2\left(K_0\right) = \frac{1}{I(I-1)} \sum_{i=1}^I\left(\hat{\tau}_i-\bar{\tau} \right)^2
,$ where $\hat{\tau}_i = \sum_{j=1}^2 (2Z_{i j}-1)(R_{ij}-K_0D_{ij})$ and $\bar{\tau} = \frac{1}{I}\sum_{i=1}^I \hat{\tau}_i$. 
Under mild regularity conditions and conditional on $\mathcal{F}, \widetilde{\mathbf{Z}}^{\text{obs}}_{\vee}, \widetilde{\mathbf{Z}}^{\text{obs}}_{\wedge}$, the test that rejects $H_0^L: \text{LB} = l$ when $ |T(l;K_0)| \geq z_{1-\alpha/2}\sqrt{S^2(K_0)}$ is an asymptotically valid level-$\alpha$ test, where $z_{1-\alpha/2}$ is the $(1-\alpha/2)$-th quantile of the standard normal distribution. 
\end{proposition}

\begin{remark}
    By replacing $K_0$ with $K_1$ in Proposition \ref{prop: inference}, a valid test can be derived analogously for testing the null hypothesis $H_0^U: \text{UB} = u$. 
    A one-sided level-$\alpha$ confidence interval of $ \text{LB}$ is
    $\left[ T(l;K_0) +l - z_{1-\alpha}\sqrt{S^2(K_0)}, \, \infty\right)$, where $z_{1-\alpha}$ is the $(1-\alpha)$-th quantile of the standard normal distribution.
    A valid confidence interval of $[\text{LB}, \text{UB}]$ may then be obtained by combining the two one-sided, level-$\alpha/2$ confidence intervals of $\text{LB}$ and $\text{UB}$.
\end{remark}

\begin{remark}
    The variance estimator $S^2(K_0)$ in general overestimates the true variance \citep{imai2008variance}. One set of sufficient conditions for it to be an unbiased estimator of the true variance is when the proportional treatment effect model $r_{Tij}-r_{Cij} = \beta (d_{Tij}-d_{Cij})$ holds \citep{small2008war}, the monotonicity assumption holds, and one of the following three conditions holds: (i) all units are compliers; (ii) each pair has
one complier and one always-taker or never-taker; (iii) no units are compliers. Such a condition appears restrictive in practice, so the variance estimator $S^2(l)$ would tend to overestimate the true variance and the test based on it would tend to be conservative.
\end{remark}

A standard approach to improving the asymptotic precision of randomization-based inference is through regression adjustment (see, e.g.,  \citealp{lin2013agnostic,fogarty2018mitigating,fogarty2018regression,li2020rerandomization, su2021model,lei2021regression,zhao2022regression}, among others). Proposition \ref{prop: reg_inference} adapts this idea to making inference of the proposed partial identification bounds by proposing a regression-assisted variance estimator for $T(l;K_0)$. 

\begin{proposition}\label{prop: reg_inference}
   Let $Q$ denote an $I\times p$ matrix with $p<I$ and $H_Q = Q(Q^TQ)^{-1}Q^T$ be the hat matrix for $Q$ with $h_{Qi}$ on the $i$-th diagonal. 
   Let $\hat{\tau}_Q$ be an $I\times1$ column vector with the $i$-th entry $\hat{\tau}_i/\sqrt{1-h_{Qi}}$,  where $\hat{\tau}_i = \sum_{j=1}^2 (2Z_{i j}-1)(R_{ij}-K_0D_{ij})$ as defined in Proposition \ref{prop: inference}. Let $\boldsymbol{\tau}$ be an $I\times1$ column vector with the $i$-th entry $\tau_i = \frac{1}{2}\sum_{j=1}^2\left\{r_{Tij}  - r_{Cij}- K_0(d_{Tij}  - d_{Cij})\right\} $.
   Define $S^2_Q(K_0) = I^{-2}\hat{\tau}^T_Q(I-H_Q)\hat{\tau}_Q$. Under conditions S1-S2
in Supplemental Material A, as $I \rightarrow \infty$, 
\begin{equation*}
    IS^2_Q(K_0)  - var\{\sqrt{I}T(l;K_0)\mid \mathcal{F}, \widetilde{\mathbf{Z}}_{\vee}, \widetilde{\mathbf{Z}}_{\wedge}\} \xrightarrow{p} \lim\limits_{I\to \infty} \frac{1}{I} \boldsymbol{\tau}^T(I-H_Q)\boldsymbol{\tau}  \geq 0.
\end{equation*}
\end{proposition}

\begin{remark}
Proposition \ref{prop: reg_inference} indicates that $IS^2_{Q}(K_0)$ is a  conservative  estimator for the variance of $\sqrt{I}T(l;K_0)$ in probability, so testing $H_0^L$ based on the statistic $T(l;K_0)/{\sqrt{S^2_{Q}(K_0)}}$ and a Gaussian reference distribution are asymptotically valid.
When $Q=\boldsymbol{e}$, a column vector of 1's, $S^2_{Q}(K_0)$ reduces to the classical variance estimator $S^2(K_0)$ as defined in Proposition \ref{prop: inference}. 
In general, for $Q = \left[\boldsymbol{e}, \boldsymbol{V}\right]$, where $\boldsymbol{V}$ is a matrix whose $i$-th row contains centered covariate $(\boldsymbol{x}_{i1}, \boldsymbol{x}_{i2})$, $IS^2_{\boldsymbol{e}}(K_0)  - IS^2_{Q}(K_0) \xrightarrow{p} \beta_V^T\Sigma_V^{-1}\beta_V \geq 0$, where $\Sigma_V = \lim\limits_{I\to \infty} I^{-1}\boldsymbol{V}^T\boldsymbol{V}$ and $\beta_V = \lim\limits_{I\to \infty} I^{-1}\boldsymbol{V}^T\boldsymbol{\tau}$.
Therefore, $IS^2_{Q}(K_0)$ is asymptotically no more conservative than $IS^2(K_0)$ if $Q$ contains any covariate information, though the degree of conservativeness reduction depends on how well the covariates predict $\hat{\tau}_i$ \citep{fogarty2018regression}. 
\end{remark}

\subsection{Inference under a IV-dose-dependent, biased randomization assumption}
\label{subsec: inference under relaxed RI}
Although matching helps remove most overt bias, residual bias due to inexact matching and from unmeasured confounders could bias the IV dose assignment within each pair and induce a so-called biased randomization scheme \citep{rosenbaum2002observational,rosenbaum2010design,fogarty2020studentized,fogarty2021biased,chen2023testingbiased}. Randomization inference under a biased randomization scheme is useful in two scenarios. First, when the minimal degree of bias due to residual confounding from observed covariates can be quantified by formal statistical tests \citep{chen2023testingbiased}, primary analyses should adopt a biased randomization scheme. Second, since the observed data contains no information about unmeasured confounding bias, a sensitivity analysis that further relaxes the randomization or biased randomization assumption in the primary analysis is always warranted. 

Recall that $\pi_{i} = P(\widetilde{Z}_{i1} = \widetilde{Z}^{\text{obs}}_{i1} \vee \widetilde{Z}^{\text{obs}}_{i2}, \widetilde{Z}_{i2} = \widetilde{Z}^{\text{obs}}_{i1}\wedge  \widetilde{Z}^{\text{obs}}_{i2}\mid \mathcal{F}, \widetilde{\mathbf{Z}}^{\text{obs}}_{\vee}, \widetilde{\mathbf{Z}}^{\text{obs}}_{\wedge})$. Following \citet{rosenbaum2023propensity}, we consider a principal unobserved covariate, $u_{ij}\in [0, 1]$, associated with each participant $ij$ and supported on the unit interval. We consider the following biased IV dose assignment model $\mathcal{M}_\gamma$ within each matched pair $i$ with an observed IV dose pair $( \widetilde{Z}^{\text{obs}}_{i1} \vee \widetilde{Z}^{\text{obs}}_{i2}, \widetilde{Z}^{\text{obs}}_{i1}\wedge  \widetilde{Z}^{\text{obs}}_{i2})$ and some $\gamma\geq 0$:
\begin{equation*}
    \begin{split}
       \pi_i = \frac{\exp\left\{\gamma\cdot\left(\widetilde{Z}^{\text{obs}}_{i1} \vee \widetilde{Z}^{\text{obs}}_{i2} - \widetilde{Z}^{\text{obs}}_{i1}\wedge  \widetilde{Z}^{\text{obs}}_{i2}\right)\cdot u_{i1}\right\}}{\exp\left\{\gamma \cdot\left(\widetilde{Z}^{\text{obs}}_{i1} \vee \widetilde{Z}^{\text{obs}}_{i2} - \widetilde{Z}^{\text{obs}}_{i1}\wedge  \widetilde{Z}^{\text{obs}}_{i2}\right)\cdot u_{i1}\right\} + \exp\left\{\gamma \cdot\left(\widetilde{Z}^{\text{obs}}_{i1} \vee \widetilde{Z}^{\text{obs}}_{i2} - \widetilde{Z}^{\text{obs}}_{i1}\wedge  \widetilde{Z}^{\text{obs}}_{i2}\right)\cdot u_{i2}\right\}},
    \end{split}
\end{equation*}
for $i = 1, \dots, n$, or equivalently,
\begin{equation*}
      \frac{1}{\exp\left\{\gamma \cdot \left(\widetilde{Z}^{\text{obs}}_{i1} \vee \widetilde{Z}^{\text{obs}}_{i2} - \widetilde{Z}^{\text{obs}}_{i1}\wedge  \widetilde{Z}^{\text{obs}}_{i2}\right)\right\} + 1} \leq \pi_i \leq  \frac{\exp\left\{\gamma \cdot \left(\widetilde{Z}^{\text{obs}}_{i1} \vee \widetilde{Z}^{\text{obs}}_{i2} - \widetilde{Z}^{\text{obs}}_{i1}\wedge  \widetilde{Z}^{\text{obs}}_{i2}\right)\right\}}{\exp\left\{\gamma \cdot \left(\widetilde{Z}^{\text{obs}}_{i1} \vee \widetilde{Z}^{\text{obs}}_{i2}- \widetilde{Z}^{\text{obs}}_{i1}\wedge  \widetilde{Z}^{\text{obs}}_{i2}\right)\right\} + 1}.
    \end{equation*} 

    \noindent The biased randomization model $\mathcal{M}_\gamma$ is analogous to that in \citet{rosenbaum1989sensitivity} and referred to as a primal sensitivity analysis \citep{gastwirth1998dual}. The model reduces to the classical Rosenbaum bounds model \citep{rosenbaum1987sensitivity,rosenbaum2002observational} when $( \widetilde{Z}^{\text{obs}}_{i1} \vee \widetilde{Z}^{\text{obs}}_{i2}, \widetilde{Z}^{\text{obs}}_{i1}\wedge  \widetilde{Z}^{\text{obs}}_{i2}) = (1, 0)$. 
    Under $\mathcal{M}_\gamma$, the degree of bias in each matched pair depends critically on the IV dose difference. For instance, if a matched pair happens to pair two participants with the same IV dose, that is, $\widetilde{Z}^{\text{obs}}_{i1} = \widetilde{Z}^{\text{obs}}_{i2}$, then $\mathcal{M}_\gamma$ would imply $\pi_{i} = 1/2$, that is, IV dose assignment becomes randomized for this pair. As the IV dose difference increases, $\mathcal{M}_\gamma$ would stipulate a larger degree of potential bias within each pair. Instead of allowing a dose-dependent maximal bias within each pair, Rosenbaum's classical $\Gamma$ sensitivity analysis model \citep{rosenbaum1987sensitivity,fogarty2021biased} places a uniform bound on the maximal biased randomization probability across all pairs, which could be unduly conservative for some matched pairs, e.g., those with $\widetilde{Z}^{\text{obs}}_{i1} = \widetilde{Z}^{\text{obs}}_{i2}$.
  
    Write $\mathcal{F}'=\{(\mathbf{x}_{ij}, u_{ij}, d_{Tij}, d_{Cij}, r_{Tij}, r_{Cij}): 0\leq {u}_{ij} \leq 1,~i=1,\dots, I,~j=1,2 \}$. Theorem \ref{thm: biased_inference} derives an asymptotically valid test for the null hypothesis $H_0^L$ under the biased randomization model $\mathcal{M}_\gamma$ for some $\gamma\geq 0$, .

\begin{theorem}\label{thm: biased_inference}
For a fixed $\gamma \geq 0$, define $\Gamma_i = 
\exp\left\{\gamma\cdot\left(\widetilde{Z}^{\text{obs}}_{i1} \vee \widetilde{Z}^{\text{obs}}_{i2} - \widetilde{Z}^{\text{obs}}_{i1}\wedge  \widetilde{Z}^{\text{obs}}_{i2}\right)\right\} 
\geq 1$ to be the IV-dose-dependent odds $\pi_i/(1 - \pi_i)$ under $\mathcal{M}_\gamma$ in matched pair $i$. For each matched pair $i$, define the following scaled treated-minus-control difference in the observed outcome:
\begin{equation}\label{eq: tau_hat under bias}
    \hat{\tau}_{i,\Gamma_i} = \frac{\Gamma_i + 1}{4\Gamma_i} \left\{ \left(\Gamma_i + 1\right)\hat{\tau}_i - \left(\Gamma_i - 1\right) \left|\hat{\tau}_i\right| \right\},
\end{equation}
where $\hat{\tau}_i$ is defined as in Proposition \ref{prop: inference}. Let $\bar{\tau}_{\gamma} = \frac{1}{I}\sum_{i=1}^I \hat{\tau}_{i,\Gamma_i}$ be the sample average across all pairs and $
S^2\left(\gamma; K_0\right) = \frac{1}{I(I-1)} \sum_{i=1}^I\left(\hat{\tau}_{i, \Gamma_i}-\bar{\tau}_{\gamma}\right)^2
$ the usual variance estimator for the sample mean. Under the biased IV assignment model $\mathcal{M}_\gamma$ and conditional on $\mathcal{F}', \widetilde{\mathbf{Z}}^{\text{obs}}_{\vee}, \widetilde{\mathbf{Z}}^{\text{obs}}_{\wedge}$, the test that rejects $H^L_0:  \text{LB} = l$ when
\begin{equation}\label{eq: test under bias}
   \bar{\tau}_{\gamma}-(l-K_0) \geq z_{1-\alpha}\sqrt{S^2\left(\gamma; K_0\right)}
\end{equation}
is an asymptotically valid level-$\alpha$ test, where $z_{1-\alpha}$ is the $(1-\alpha)$-th quantile of the standard normal distribution. A valid one-sided, level-$\alpha$ confidence interval can then be constructed by inverting the hypothesis testing.
\end{theorem}

\begin{remark}\label{remark: gamma = 0 reduce to RI}
When $\gamma = 0$, $\mathcal{M}_\gamma$ reduces to the the randomization scheme studied in Section \ref{subsec: inference under RI}. In this case, $\Gamma_i = 1$ for all $i$, and $\hat{\tau}_{i, \Gamma_i}$ defined in \eqref{eq: tau_hat under bias} reduces to the usual treated-minus-control difference in the observed outcome in each pair. The test \eqref{eq: test under bias} and the associated confidence interval reduce to those studied in Section \ref{subsec: inference under RI} under the randomization assumption.
\end{remark}

\begin{remark}
    By replacing $K_0$ with $K_1$ in Theorem \ref{thm: biased_inference}, a valid test can be derived analogously for testing the null hypothesis $H_0^U: \text{UB} = u$. 
    A valid level-$\alpha$ confidence interval of the partial identification bound $[\text{LB}, \text{UB}]$ may then be obtained by combining two one-sided, level-$\alpha/2$ confidence intervals of $\text{LB}$ and $\text{UB}$.
\end{remark}

The test derived in Theorem \ref{thm: biased_inference} also applies to testing the effect ratio estimand under $\mathcal{M}_\gamma$, as formalized in Proposition \ref{prop: biased_inference for lambda}.

\begin{proposition}\label{prop: biased_inference for lambda}
Consider testing the null hypothesis $H^{\lambda}_0:  \lambda = \lambda_0$ regarding the effect ratio under the biased IV assignment model $\mathcal{M}_\gamma$. Let $\hat{\tau}_i = \sum_{j=1}^2 (2Z_{i j}-1)(R_{ij}-\lambda_0D_{ij})$. Define $\hat{\tau}_{i,\Gamma_i}$, $\bar{\tau}_{\gamma}$, and $S^2\left(\gamma; \lambda_0\right)$ as in Theorem \ref{thm: biased_inference}. Conditional on $\mathcal{F}'$, $\widetilde{\mathbf{Z}}^{\text{obs}}_{\vee}$, and $\widetilde{\mathbf{Z}}^{\text{obs}}_{\wedge}$, the test that rejects $H^{\lambda}_0$ when
$
   \bar{\tau}_{\gamma} \geq z_{1-\alpha}\sqrt{S^2\left(\gamma; \lambda_0\right)}
$
is an asymptotically valid level-$\alpha$ test. 
\end{proposition}


%% file: 6_Simulation.tex
\section{Simulation studies}
\label{sec: simulation}

We have two goals in the simulation studies. In Section \ref{subsec: simulation 1}, we illustrate how strengthening an IV in the design stage helps improve the compliance rate and thus narrow the partial identification bounds and deliver more informative inferential targets. In Section \ref{subsec: simulation 2}, we generate synthetic data modeled after the real data from the case study and verify the level of the proposed tests. 

\subsection{Simulation study I: study design and its effect on the causal estimand}
\label{subsec: simulation 1}

We generate potential outcomes data for each participant $n \in \{1, \dots, N\}$. We consider a 5-dimensional covariate as follows: $X_{n, 1} \sim \mathcal{N}(0,1), \, X_{n, 2} \sim \mathcal{N}(2,5), \,
X_{n, 3} \sim \text{Unif}[1,3], \, 
X_{n, 4} \sim \text{Unif}[-2, 0]$ and $X_{n, 5} \sim \text{Bernoulli}(0.5)$. Each participant is associated with a step function $D_n(Z = z)$ with a jump at a randomly sampled threshold $T_n \sim \text{Unif}[20,30]$. The step function $D_n(Z = z)$ describes the potential treatment received for each participant as a function of the IV dose $Z$. According to this data-generating process, each participant has one's own incentive structure. Factor $1$ specifies the distribution of the observed IV dose $Z^{obs}_{n}$ for participant $n$:
\begin{description}
    \item[Factor 1:] We consider two scenarios for $\widetilde{Z}^{obs}_{n}$: (1) $\widetilde{Z}^{obs}_{n} \sim \text{Unif}[5, 50]$ is randomized; and (2) $\widetilde{Z}^{obs}_{n} = 4X_{n, 2} + 6X_{n, 3} + \varphi_{n}$ is correlated with the observed covariates, where $\varphi_{n} \sim \text{Unif}[0, 2]$.
\end{description}
The observed treatment received $D^{obs}_{n}$ is then determined by $D^{obs}_{n} = D_{n}(\widetilde{Z}^{obs}_{n}) = \mathbb{I}(\widetilde{Z}_{n}^{obs} > T_n).$ Using the covariates and the observed IV dose $\widetilde{Z}^{obs}_{n},$ we perform non-bipartite matching. Factor 2 specifies three matching algorithms under consideration:

\begin{description}
    \item[Factor 2:] $\widetilde{\mathcal{M}}_1, \, \widetilde{\mathcal{M}}_2$ and $\widetilde{\mathcal{M}}_3$ refer to matching designs with no caliper, a small caliper of $7$ minutes, and a large caliper of $15$ minutes on the absolute difference between two within-matched-pair observed IV doses, respectively.
\end{description}
Each matching design yields $I = N/2$ matched pairs of participants, where each matched pair is indexed by $i \in [I]$ and each within-matched-pair participant indexed by $j \in \{1,2\}.$ The two observed IV doses within each matched pair, $\widetilde{Z}^{obs}_{i1} \vee \widetilde{Z}^{obs}_{i2}$ and $\widetilde{Z}^{obs}_{i1} \wedge \widetilde{Z}^{obs}_{i2},$ are fixed in each matched design. We then determine participant $ij$'s compliance status $S_{ij}$ as follows:
\begin{align*}
\text{If } D_{ij}(\widetilde{Z}^{obs}_{i1} \wedge \widetilde{Z}^{obs}_{i2}) &= D_{ij}(\widetilde{Z}^{obs}_{i1} \vee \widetilde{Z}^{obs}_{i2}) = 0, \text{ then } S_{ij} = (0,0); \\
\text{If }D_{ij}(\widetilde{Z}^{obs}_{i1} \wedge \widetilde{Z}^{obs}_{i2}) &= 0 \text{ and } D_{ij}(\widetilde{Z}^{obs}_{i1} \vee \widetilde{Z}^{obs}_{i2}) = 1, \text{ then } S_{ij} = (1,0); \\
\text{If }D_{ij}(\widetilde{Z}^{obs}_{i1} \wedge \widetilde{Z}^{obs}_{i2}) &= D_{ij}(\widetilde{Z}^{obs}_{i1} \vee \widetilde{Z}^{obs}_{i2}) = 1, \text{ then }S_{ij} = (1,1),
\end{align*}
where $S_{ij} \in \{(0,0), (1,0), (1,1)\}$ indicates that participant $ij$ is a never-taker, complier or always-taker with respect to the two observed IV doses in matched pair $i$, respectively. 

Finally, we generate the potential outcomes $r_{d = 0, ij} \sim \mathcal{N}(0,1)$. Factor 3 specifies the unit-level treatment effect and hence the data-generating process for $r_{d = 1, ij}$: 

\begin{description}
    \item[Factor 3:] We consider two scenarios for $r_{d = 1, ij}$: (1) $r_{d=1,ij} = r_{d=0,ij} + \psi_{ij}$, where the unit-level treatment effect $\psi_{ij}$ is independent of participant $ij$'s compliance status and follows $\psi_{ij} \sim \text{Unif}[4,6]$; and (2) $r_{d=1, ij} = r_{d=0,ij} + \phi_{ij},$ where the unit-level treatment effect $\phi_{ij}$ depends on participant $ij$'s compliance status as follows: $\phi_{ij} \sim \text{Unif}[2,5]$ for a complier, $\phi_{ij} \sim \text{Unif}[4,6]$ for an always-taker, and $\phi_{ij} \sim \text{Unif}[1,3]$ for a never-taker.
\end{description}
Participant $ij$'s potential outcomes $r_{T_{ij}}$ and $r_{C_{ij}}$ are then determined by the compliance status $S_{ij}$ and the potential outcomes $(r_{d=0, ij}, r_{d=1,ij})$ as follows: $(r_{T_{ij}}, r_{C_{ij}})$ equals $(r_{d=0,ij}, r_{d=0,ij})$ for never-takers, $(r_{d=1,ij}, r_{d=0,ij})$ for compliers, and $(r_{d=1,ij}, r_{d=1,ij})$ for always-takers. For each matched dataset, we calculate $LB$ and $UB$ as described in Section \ref{subsec: inference} with  $(K_0, K_1) \in \{(4,6), (1,6)\}$ for Scenario 1 or 2 in Factor 3, respectively. We generate data for $N = 1,000$ participants and repeat the simulation $1,000$ times under each data-generating process. 


\begin{table}[!ht]
\centering
\caption{Average widths of the partial identification bounds across $1,000$ simulations. In the first column, $F_{ij}$ refers to the $j$-th scenario in the $i$-th factor.}
\label{tb: simulation 1}
\begin{tabular}{rrrr}
  \hline
 & $\widetilde{\mathcal{M}}_1$ & $\widetilde{\mathcal{M}}_2$ & $\widetilde{\mathcal{M}}_3$ \\ 
  \hline
$F_{11} \times F_{31}$ & 1.03 & 0.72 & 0.41 \\ 
  $F_{11} \times F_{32}$  & 2.57 & 1.80 & 1.02 \\ 
  $F_{12} \times F_{31}$ & 1.83 & 1.33 & 0.83 \\ 
  $F_{12} \times F_{32}$ & 4.58 & 3.34 & 2.07 \\ 
   \hline
\end{tabular}
\end{table}

Table \ref{tb: simulation 1} summarizes the empirical means of the widths of partial identification bounds, $\Delta = UB - LB$, across $1,000$ simulations for each combination of the three factors. In the first column of Table \ref{tb: simulation 1}, $F_{ij}$ denotes the $j$-th scenario in the $i$-th factor; for instance, $F_{11} \times F_{32}$ denotes the simulation setting of randomized IVs $\widetilde{Z}_k^{obs}$ and potential outcomes $r_{d=1,k}$ that depend on the compliance status. Across the four data-generating processes specified by Factors $1$ and $3$, the more we strengthen the continuous IV, the narrower the widths of the partial identification bounds, yielding more useful information on the SATE. This is in line with our expectation, as the width of a partial identification bound is proportional to the compliance rate, which increases with the IV dose caliper.

Table S1 in Supplemental Material B further summarizes within-matched-pair covariate balance under each data-generating process using the average absolute value of the standardized mean difference (SMD) across $1,000$ simulations. The average absolute value of the SMD increases with the IV dose caliper, with $X_2$ and $X_3$ exhibiting the largest increase when they are correlated with the IV dose as specified by Factor 1 Scenario 2. Nevertheless, SMDs are almost always less than $0.1$, which usually indicates a good covariate balance in empirical studies \citep{silber2001matching}.

\subsection{Simulation study II: validity of the proposed tests}
\label{subsec: simulation 2}

In this simulation study, we verify the validity of the proposed statistical tests under randomization inference (Proposition \ref{prop: inference}) and biased randomization inference (Theorem \ref{thm: biased_inference}). For every unit $j$ in matched pair $i$, we first generate potential outcomes data. Specifically, we sample two continuous IV doses $\widetilde{Z}^{obs}_{i1} \wedge \widetilde{Z}^{obs}_{i2}$ and $\widetilde{Z}^{obs}_{i1} \vee \widetilde{Z}^{obs}_{i2}$ with replacement from the real NICU dataset. Unit $ij$'s compliance status $S_{ij}(\widetilde{Z}^{obs}_{i1} \wedge \widetilde{Z}^{obs}_{i2}, \widetilde{Z}^{obs}_{i1} \vee \widetilde{Z}^{obs}_{i2}) \in \{(1,0), (1,1), (0,0)\}$ is generated as follows:
\begin{equation*}
\begin{split}
    &P\{S_{ij} = (1,0)\} = C^{-1}; \\
    &P\{S_{ij} = (1,1)\} = C^{-1}\cdot\exp \{-0.2\times (\widetilde{Z}^{obs}_{i1} \vee \widetilde{Z}^{obs}_{i2} - \widetilde{Z}^{obs}_{i1} \wedge \widetilde{Z}^{obs}_{i2}) \};\\
    &P\{S_{ij} = (0,0)\} = C^{-1} \cdot \exp \{ -\alpha_i (\widetilde{Z}^{obs}_{i1} \vee \widetilde{Z}^{obs}_{i2}) \},
\end{split}
\end{equation*}
where $C$ is a normalizing constant, and $ij$'s compliance status depends on the two IV doses in each matched pair. The latent compliance category $S_{ij}$ then determines $(d_{Tij}, d_{Cij})$ as defined in \eqref{eq: definition of dt and dc}.  

We consider two scenarios for each of the two types of potential outcomes $(r_{d = 0, ij}, r_{d = 1, ij})$:
\begin{description}
    \item[Continuous case:] \textsf{Scenario 1}: $r_{d=0, ij} \sim \mathcal{N}(0,1)$ and $r_{d=1, ij} = r_{d=0, ij}$; \textsf{Scenario 2}: $r_{d=0, ij} \sim \mathcal{N}(0,1)$ and $r_{d=1, ij} = r_{d=0, ij} + \tau_{ij},$ where $\tau_{ij} \sim \text{Unif}[-1,1]$.
    
    \item[Binary case:] \textsf{Scenario 1}: $P(r_{d=0, ij} = 1) = \text{expit}(0.5),$ where $\text{expit}(x) = \exp(x)/(1+\exp(x))$ is the inverse of the logit function, and $r_{d=1,ij} = r_{d=0,ij}$; \textsf{Scenario 2}: $P(r_{d=0, ij} = 1) = \text{expit}(0.5)$ and $P(r_{d=1, ij} = 1) = \text{expit}(0.5 + \tau_{ij}),$ where $\tau_{ij} \sim \text{Unif}[-1,1]$.
\end{description}

\noindent For both outcome types, \textsf{Scenario 1} represents a sharp null hypothesis of no unit-level treatment effect, while \textsf{Scenario 2} allows a heterogeneous treatment effect. Once we generate the potential outcomes data, for every matched pair $i$, we let the participant with the larger potential outcome under control, i.e., $r_{d=0,ij}$, be more likely to receive the treatment under a biased IV dose assignment mechanism $\pi_i,$ which is introduced below. In this way, the IV doses are correlated with both the treatment assignment and the potential outcomes $(r_{d=0,ij}, r_{d=1,ij})$. 

According to this data-generating process, unit $ij$'s potential outcomes $(r_{T_{ij}}, r_{C_{ij}})$ are completely specified by the compliance category $S_{ij}$ and potential outcomes $(r_{d = 1, ij}, r_{d = 0, ij})$. Finally, we generate the observed IV dose $\widetilde{Z}_{ij}^{obs}$ according to one of the following IV dose assignment mechanisms: 
\begin{equation*}
    \begin{split}
        \pi_{i} &= P(\widetilde{Z}_{i1} = \widetilde{Z}^{\text{obs}}_{i1} \vee \widetilde{Z}^{\text{obs}}_{i2}, \widetilde{Z}_{i2} = \widetilde{Z}^{\text{obs}}_{i1}\wedge  \widetilde{Z}^{\text{obs}}_{i2}\mid \mathcal{F}', \widetilde{\mathbf{Z}}^{\text{obs}}_{\vee}, \widetilde{\mathbf{Z}}^{\text{obs}}_{\wedge}) \\
&\sim \text{Unif} \left[\frac{1}{2}, \frac{\exp\left\{\gamma (\widetilde{Z}^{\text{obs}}_{i1} \vee \widetilde{Z}^{\text{obs}}_{i2} - \widetilde{Z}^{\text{obs}}_{i1} \wedge \widetilde{Z}^{\text{obs}}_{i2})\right\}}{\exp\left\{\gamma (\widetilde{Z}^{\text{obs}}_{i1} \vee \widetilde{Z}^{\text{obs}}_{i2} - \widetilde{Z}^{\text{obs}}_{i1} \wedge \widetilde{Z}^{\text{obs}}_{i2})\right\} + 1} \right]~(\textbf{Mechanism I}),~\text{or} \\
&= \frac{\exp\left\{\gamma (\widetilde{Z}^{\text{obs}}_{i1} \vee \widetilde{Z}^{\text{obs}}_{i2} - \widetilde{Z}^{\text{obs}}_{i1} \wedge \widetilde{Z}^{\text{obs}}_{i2})\right\}}{\exp\left\{\gamma (\widetilde{Z}^{\text{obs}}_{i1} \vee \widetilde{Z}^{\text{obs}}_{i2} - \widetilde{Z}^{\text{obs}}_{i1} \wedge \widetilde{Z}^{\text{obs}}_{i2})\right\} + 1}~(\textbf{Mechanism II}).
    \end{split}
\end{equation*}

The observed data comprises the observed IV dose $\widetilde{Z}_{ij}^{obs},$ treatment $D_{ij},$ and outcome $R_{ij}$, with the observed $D_{ij}$ and $R_{ij}$ completely determined by $\widetilde{Z}_{ij}^{obs}$ and the potential outcomes data. We vary the following factors in the simulation study: 
\begin{description}
    \item[Factor 1:] Number of matched pairs $I$: $100$, $500$, $1000$ and $2000$.
    \item[Factor 2:] IV dose assignment probability $\gamma$: $0$, $0.025$ and $0.05$. If $\gamma = 0$, the two IV doses are randomly assigned within each pair and the randomization assumption holds in either mechanism. Otherwise, the IV dose assignment could be biased within each pair and the degree of bias depends on the difference between the two IV doses.
\end{description}
In \textsf{Scenario 1}, we have $(K_0, K_1) = (0, 0)$ regardless of whether the outcome is continuous or binary. In \textsf{Scenario 2}, $(K_0, K_1) = (-1,1)$ for both the continuous and binary potential outcomes. In each setting, we repeat the simulation $1,000$ times. For each simulated dataset, we calculate the ground truth $LB_0 = \frac{1}{N} \sum_{ij} (r_{Tij} - r_{Cij}) - K_0 \times \frac{1}{N} \sum_{ij} (d_{Tij} - d_{Cij}) + K_0$ and construct the 95\% confidence interval for $LB$ according to Proposition \ref{prop: inference} and Theorem \ref{thm: biased_inference}. We apply Proposition \ref{prop: inference} and Theorem \ref{thm: biased_inference} to data generated under the biased IV dose assignment scheme, i.e., $\gamma \neq 0$, to illustrate that the 95\% confidence intervals constructed according to Theorem \ref{thm: biased_inference} have the desired level while those constructed using Proposition \ref{prop: inference} may not. 

Table \ref{tb: simulation 2} considers the case where the potential outcomes $(r_{d=0, ij}, r_{d=1, ij})$ are continuous and summarizes the coverage of $95\%$ confidence intervals under Proposition \ref{prop: inference} and Theorem \ref{thm: biased_inference} across $1,000$ simulations in \textsf{Scenario 1} and \textsf{Scenario 2}. In the absence of a biased IV dose assignment, i.e., $\gamma = 0$, the two IV assignment mechanisms collapse to one, and the test statistic almost or fully achieves nominal coverage in both scenarios across a range of sample sizes. 

As the degree of bias in the IV dose assignment increases in either scenario, the test exhibits exceedingly small or zero coverage under Proposition \ref{prop: inference}. In such cases, conducting randomization inference for $LB$ overestimates $LB$ and leads to under-coverage, even though the test tends to be conservative in the first place. By contrast, the test constructed according to Theorem \ref{thm: biased_inference} tends to achieve perfect coverage under \textbf{Mechanism I} and close to nominal coverage under \textbf{Mechanism II}, whereby the degree of coverage varies with unit-level treatment effect heterogeneity.

Two sources of conservativeness account for this trend. First, the test constructed according to Theorem \ref{thm: biased_inference} assumes a worst-case allocation of bias within each pair; hence, the test will tend to be more conservative for \textbf{Mechanism 1} compared to \textbf{Mechanism 2}. Second, even under the worst-case bias allocation in \textbf{Mechanism 2}, the variance estimator of the test statistic is a conservative estimator of the true variance in \textsf{Scenario 2}; therefore, the test still tends to be conservative with a coverage often exceeding its nominal level. We observe similar trends for the case where the potential outcomes $(r_{d=0, ij}, r_{d=1, ij})$ are binary; see Supplemental Material B for details.

\begin{table}[!ht]
\centering
\caption{Simulation results when the potential outcomes $(r_{d=0, ij}, r_{d=1, ij})$ are continuous under two scenarios: causal null hypothesis (\textsf{Scenario 1}) and heterogeneous unit-level treatment effect (\textsf{Scenario 2}). We report the average coverage of $95\%$ confidence intervals under Proposition \ref{prop: inference} and Theorem \ref{thm: biased_inference} across $1,000$ simulations.}
\scalebox{0.85}{
\label{tb: simulation 2}
\begin{tabular}{lrrcccc}
 \toprule
& & & \multicolumn{2}{c}{\textsf{Scenario 1}} & \multicolumn{2}{c}{\textsf{Scenario 2}} \\ \cmidrule(lr){4-5} \cmidrule(lr){6-7} 
& & &95\% CI Coverage & 95\% CI Coverage & 95\% CI Coverage & 95\% CI Coverage \\
 & $I$ & $\gamma$ & Proposition 1 & Proposition 3 & Proposition 1 & Proposition 3 \\
 \hline
Mechanism I & 100 & 0.000 & 0.937 & 0.946 & 0.953 & 0.944 \\ 
&  500 & 0.000 & 0.949 & 0.951 & 0.961 & 0.958 \\ 
&  1000 & 0.000 & 0.955 & 0.945 & 0.972 & 0.971 \\ 
&  2000 & 0.000 & 0.954 & 0.941 & 0.965 & 0.959 \\ 
&  100 & 0.025 & 0.797 & 0.997 & 0.826 & 0.998 \\ 
&  500 & 0.025 & 0.349 & 1.000 & 0.416 & 1.000 \\ 
&  1000 & 0.025 & 0.100 & 1.000 & 0.142 & 1.000 \\ 
&  2000 & 0.025 & 0.002 & 1.000 & 0.005 & 1.000 \\ 
&  100 & 0.050 & 0.537 & 1.000 & 0.611 & 1.000 \\ 
&  500 & 0.050 & 0.020 & 1.000 & 0.030 & 1.000 \\ 
&  1000 & 0.050 & 0.000 & 1.000 & 0.000 & 1.000 \\ 
&  2000 & 0.050 & 0.000 & 1.000 & 0.000 & 1.000 \\ 
   \hline
Mechanism II & 100 & 0.000 & 0.953 & 0.951 & 0.963 & 0.966 \\ 
&  500 & 0.000 & 0.953 & 0.945 & 0.960 & 0.954 \\ 
&  1000 & 0.000 & 0.945 & 0.963 & 0.961 & 0.964 \\ 
&  2000 & 0.000 & 0.947 & 0.951 & 0.963 & 0.957 \\ 
&  100 & 0.025 & 0.405 & 0.932 & 0.522 & 0.973 \\ 
&  500 & 0.025 & 0.002 & 0.943 & 0.007 & 0.996 \\ 
&  1000 & 0.025 & 0.000 & 0.946 & 0.000 & 0.998 \\ 
&  2000 & 0.025 & 0.000 & 0.940 & 0.000 & 1.000 \\ 
&  100 & 0.050 & 0.028 & 0.893 & 0.069 & 0.980 \\ 
&  500 & 0.050 & 0.000 & 0.922 & 0.000 & 0.998 \\ 
&  1000 & 0.050 & 0.000 & 0.920 & 0.000 & 1.000 \\ 
&  2000 & 0.050 & 0.000 & 0.944 & 0.000 & 1.000 \\ 
\bottomrule
\end{tabular}
}
\end{table}

%% file: 7_Case_study.tex
\section{Revisiting the NICU study}
\label{sec: case study}

\subsection{Assessing the randomization assumption}
\label{subsec: assess RA}
An IV-based matched cohort design seeks to embed the retrospective observational data into an ``approximate" randomized encouragement experiment. As a first step towards a randomization-based outcome analysis for the design \textsf{M1}, informal and formal balance diagnostics are often used to justify the \emph{randomization assumption}, which states that the two IV doses within each matched pair are randomly assigned \citep{silber2001matching,gagnon2019cpt,yu2020evaluating}. If the randomization assumption is rejected, then some relaxed version of it, such as the biased randomization assumption, may be tested and a minimal degree of residual imbalance due to inexact matching on the observed covariates may be quantified \citep{chen2023testingbiased}. If there is strong evidence against the randomization assumption, then one could opt to conduct the primary analysis under a biased randomization scheme. Regardless of whether the randomization or biased randomization assumption is used in the primary analysis, a sensitivity analysis that further examines the impact of unmeasured IV-outcome confounding is recommended.

\begin{figure}%
    \centering
    \subfloat[\centering   
    \label{fig:CPT null distribution}]
{{\includegraphics[width=0.485\textwidth]{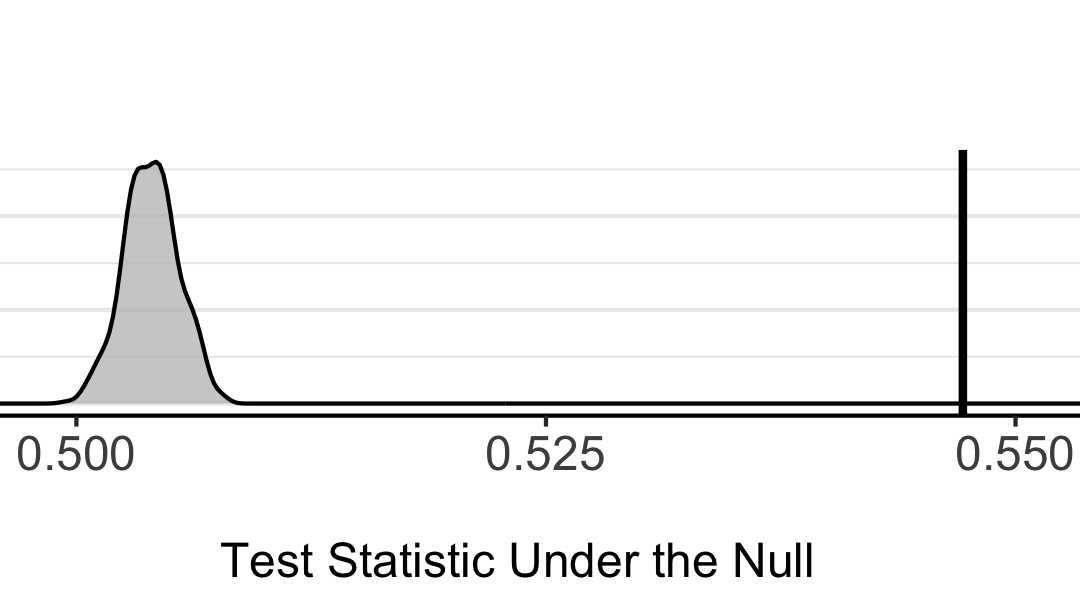} }}%
    \qquad
    \subfloat[\centering  \label{fig:Gamma distribution}]{{\includegraphics[width=0.45\textwidth]{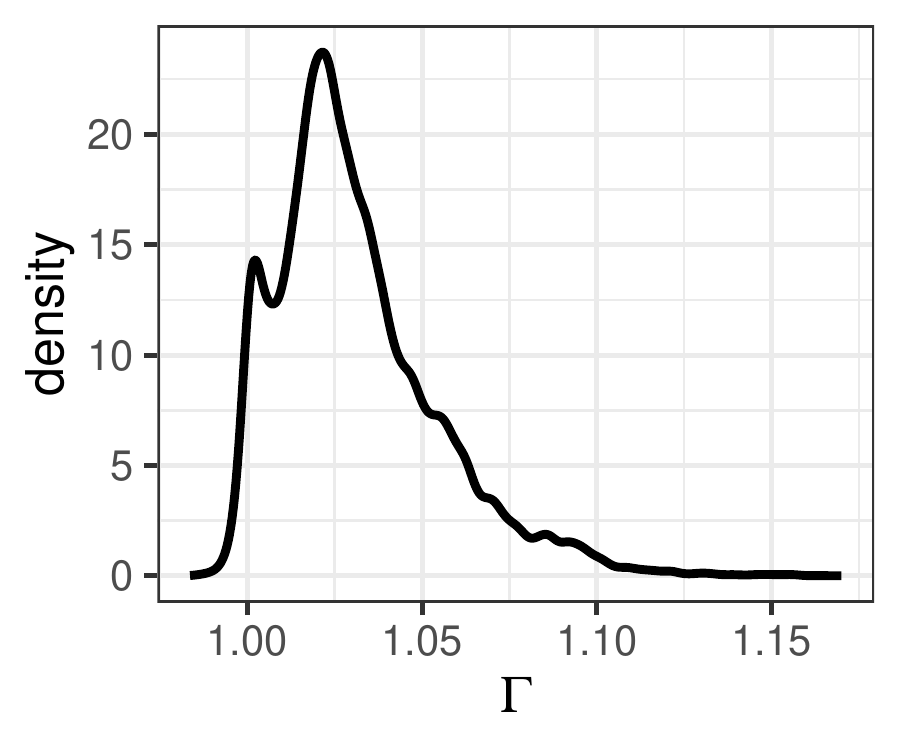} }}%
    \caption{Panel (a): Distribution of the in-sample classification accuracy rate under the null hypothesis for \textsf{M1}. The vertical bar denotes the observed test statistic. Panel (b): density plot of $\Gamma_i$'s with $\gamma = 0.0012$ in the design $\textsf{M1}$.
    }%
    \label{fig: gamma}
\end{figure}

We assessed the randomization assumption using the Classification Permutation Test (CPT) \citep{gagnon2019cpt}. Specifically, we tested the null hypothesis that the excess travel times of mothers within each matched pair are randomly assigned. When implementing the CPT, we fit a logistic regression model to predict which mother within each matched pair had a larger excess travel time based on the observed covariates and used in-sample classification accuracy rate as our test statistic. We then obtained the permutation-based null distribution of the test statistic  based on $500$ permutations of the IV doses. Finally, we computed an exact p-value by comparing the observed test statistic to the reference distribution. Figure \ref{fig:CPT null distribution} shows this null distribution with the observed test statistic superimposed. A large value of the test statistic indicates that the covariates, even after matching, are still predictive of the IV dose magnitude within each matched pair and serves as evidence against the randomization assumption. Thus, the randomization assumption for the design \textsf{M1} was rejected (p-value $<0.001$). 
 

To what extent was the randomization assumption violated? Towards understanding this question, we proceeded to test the following biased randomization assumption: $H_{0, \Gamma}: \frac{1}{1+\Gamma} \leq \pi_i \leq \frac{\Gamma}{1+\Gamma}$ for all $i$ \citep{chen2023testingbiased}. The null hypothesis $H_{0, \Gamma}$ implies that the IV-dose-dependent odds ratio $\Gamma_i$ for every matched pair $i$ within \textsf{M1} is uniformly bounded between $1/\Gamma$ and $\Gamma$ and hence the IV assignment probability $\pi_i$ is uniformly bounded between $1/(1+\Gamma)$ and $\Gamma/(1+\Gamma)$. Using \citeauthor{chen2023testingbiased}'s \citeyearpar{chen2023testingbiased} sample-splitting CPT, we rejected the null hypothesis $H_{0, \Gamma}$ for $\Gamma$ as large as $1.17$; therefore, we have evidence that at least one matched pair exhibits bias in the IV dose assignment odds $\Gamma_i$ at a magnitude larger than $1.17$. This implies that the smallest $\gamma$ in $\mathcal{M}_\gamma$ not falsified by the observed data and the diagnostic test equals $\gamma = \log(1.17)/(\max_i \widetilde{Z}^{\text{obs}}_{i1} \vee \widetilde{Z}^{\text{obs}}_{i2} - \widetilde{Z}^{\text{obs}}_{i1}\wedge  \widetilde{Z}^{\text{obs}}_{i2}) = \log(1.17) / 129 = 0.0012$ for \textsf{M1}. Figure \ref{fig:Gamma distribution} displays the distribution of $\Gamma_i$'s across all matched pairs in $\textsf{M1}$ when $\gamma = 0.0012$. This value of $\gamma$ and its associated sensitivity analysis model $\mathcal{M}_\gamma$ served as the basis for our biased-randomization-based primary outcome analysis.

\subsection{Primary analysis of the matched sample}
\label{subsec: case study outcome analysis}
We conducted a biased-randomization-based primary analysis for the matched sample $\textsf{M1}$ under $\mathcal{M}_\gamma$ with $\gamma = 0.0012$. We considered two outcomes of interest. The primary outcome of interest was infant mortality rate. We also considered a composite outcome that combines infant mortality rate and length of stay in the intensive care unit (ICU) as an overall measurement of health care quality; such a composite outcome is referred to as a ``placement of death" approach \citep{lin2017placement} and used in many clinical settings (see, e.g., \citealt{courtright2024default}). Specifically, this composite outcome equals the length of stay in the ICU (LOSI) if the baby survived and the top 1\% quantile of survivor LOSI if the baby died. Table \ref{table: primary analysis} summarizes
the sample size, compliance rate, mortality, LOSI among survivors, and the composite outcome in the near group (mothers with the smaller excess travel time in each pair) and far group (mothers with the larger excess travel time in each pair). 

\begin{table}[!ht]
\centering
\caption{Level of encouragement, proportion attending a high-level NICU, and mortality in the near (mothers with the smaller excess travel time in each pair) group and far (mothers with the larger excess travel time in each pair) group. }
\label{table: primary analysis}
\scalebox{0.75}{
\begin{tabular}{llcccccc} \toprule
 & & \multicolumn{2}{c}{\textsf{M1}} & \multicolumn{2}{c}{Inferential Results}
\\ 
\cmidrule(lr){3-4} \cmidrule(lr){5-6} 
 & & Near Group  & Far Group & Pt. Est. & 95\% CI \\
\midrule Primary analysis: & Sample size & 81,766 & 81,766 \\
~~Entire cohort& Excess travel time (min) & 2.07 & 27.1   \\
 &High-level NICU (\%)  & 71.9 & 35.5   \\
 &Death (\%) & 2.1 & 2.6   &  &  &  \\
 &~~Effect ratio, $\lambda$ & & &1.28 & [0.69, 1.90] \\
 &~~SATE, $\kappa$ & & & &Fig 3(a)\\
 
 &LOSI among survivors (days) & 8.16 & 7.65 &  &  \\
  &Composite outcome  & 9.64 & 9.50\\ 
  &~~Effect ratio, $\lambda$ & & & -0.38 &[-1.34, 0.59]   \\
   &~~SATE, $\kappa$ & & & &Fig 3(b)\\
\midrule Subgroup Analysis I: & Sample size & 13,454 & 13,454 \\
~~Black & Excess travel time (min) & -0.39 & 9.25   \\
 &High-level NICU (\%)  & 82.96 & 64.64   \\
 &Death (\%) & 3.54 & 3.84  \\

&~~Effect ratio, $\lambda$ & & &1.62 & [-1.16, 4.50] \\
 &~~SATE, $\kappa$ & & & &Fig 4(a) \\
 
 &LOSI among survivors (days) & 9.95 & 9.58 &  &  \\
  &Composite outcome  & 12.81 & 12.72\\ 
  &~~Effect ratio, $\lambda$ & & & -0.48 &[-1.34, 0.59]   \\
   &~~SATE, $\kappa$ & & & &Fig S1 \\
 
    
\midrule Subgroup Analysis II: & Sample size & 6,863 & 6,863 \\
~~Non-Black, Age $\geq 35$, & Excess travel time (min) & 1.95 & 26.01   \\
~~Gestational age $\leq 36$ &High-level NICU (\%)  & 77.78 & 45.32   \\
 &Death (\%) & 2.77 & 3.77   \\
&~~Effect ratio, $\lambda$ & & &3.10 & [0.92, 5.42] \\
 &~~SATE, $\kappa$ & & & &Fig 4(b) \\
 
 &LOSI among survivors (days) & 11.84 & 11.36 &  &  \\
  &Composite outcome & 14.15 & 14.58\\ 
  &~~Effect ratio, $\lambda$ & & & 1.32 &[-1.96, 4.74]   \\
   &~~SATE, $\kappa$ & & & & Fig S1 \\
\midrule Subgroup Analysis III: & Sample size & 32,294 & 32,294 \\
~~Non-Black, Age $\leq 35$, & Excess travel time (min) & 2.54 & 30.86   \\
~~Gestational age $\leq 36$ &High-level NICU (\%)  & 71.29 & 37.40   \\
 &Death (\%) & 2.31 & 2.85   \\
&~~Effect ratio, $\lambda$ & & & 2.02 & [0.85, 3.26] \\
 &~~SATE, $\kappa$ & & & &Fig 4(b) \\
 
 &LOSI among survivors (days) & 11.58 & 10.66 &  &  \\
  &Composite outcome & 13.42 & 13.72\\ 
  &~~Effect ratio, $\lambda$ & & & -0.81 &[-2.78, 1.21]   \\
   &~~SATE, $\kappa$ & & & & Fig S1 \\
\midrule Subgroup Analysis IV: & Sample size & 29,155 & 29,155 \\
~~Gestational age $>$ 36 & Excess travel time (min) & 2.70 & 31.43   \\
&High-level NICU (\%)  & 65.87 & 17.73   \\
 &Death (\%) & 0.23 & 0.41   \\
&~~Effect ratio, $\lambda$ & & & 0.35 & [0.10, 0.61] \\
 &~~SATE, $\kappa$ & & & &Fig 4(b) \\
 
 &LOSI among survivors (days) & 2.72 & 2.60 &  &  \\
  &Composite outcome & 2.73 & 2.62 \\ 
  &~~Effect ratio, $\lambda$ & & & -0.24 &[-0.42, -0.07]   \\
   &~~SATE, $\kappa$ & & & & Fig S1 \\
\bottomrule
\end{tabular}
}
\end{table}

We first made inference for the effect ratio estimand under $\mathcal{M}_\gamma$ with $\gamma = 0.0012$ using Proposition \ref{prop: biased_inference for lambda}. The effect ratio of the mortality outcome (low-level NICU vs. high-level NICU) was estimated to be $1.28\%$ with a $95\%$ confidence interval of $[0.69\%, 1.90\%]$, providing evidence that delivering at a low-level NICU increased infants' mortality rate among the $71.9\% - 35.5\% = 36.4\%$ compliers in $\textsf{M1}$. Such an estimated treatment effect on compliers is substantial, almost half of the overall mortality rate in the entire study cohort. On the other hand, the effect on the composite outcome was estimated to be $-0.38$ among compliers with a $95\%$ confidence interval of $[-1.34, 0.59]$, not providing evidence for an effect on the composite outcome. 

\begin{figure}%
    \centering
    \subfloat[\centering   Mortality rate 
    \label{fig:Death_CI_plot}]
{{\includegraphics[width=0.45\textwidth]{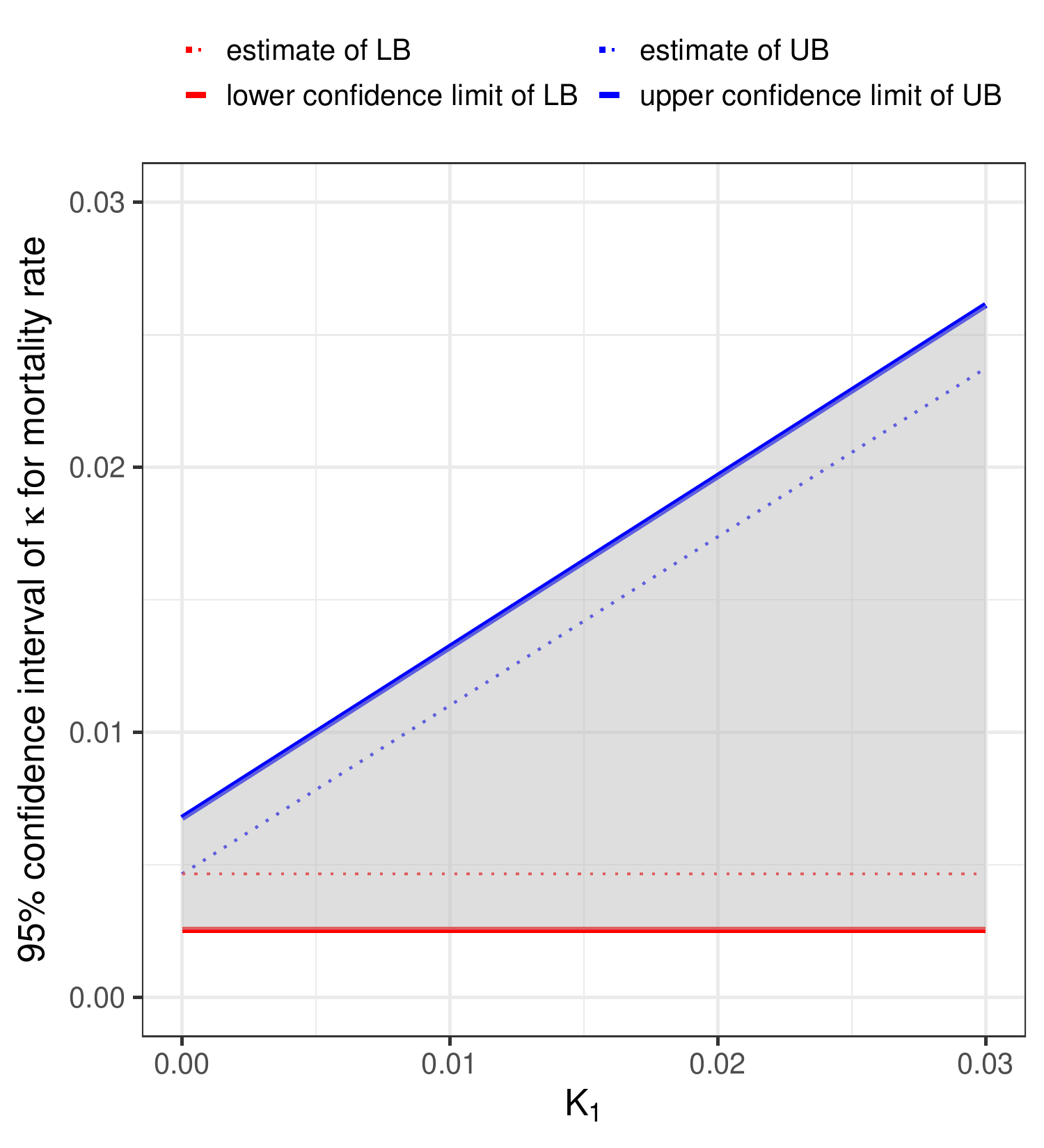} }}%
    \qquad
    \subfloat[\centering  Composite outcome\label{fig:Composite_CI_plot}]{{\includegraphics[width=0.45\textwidth]{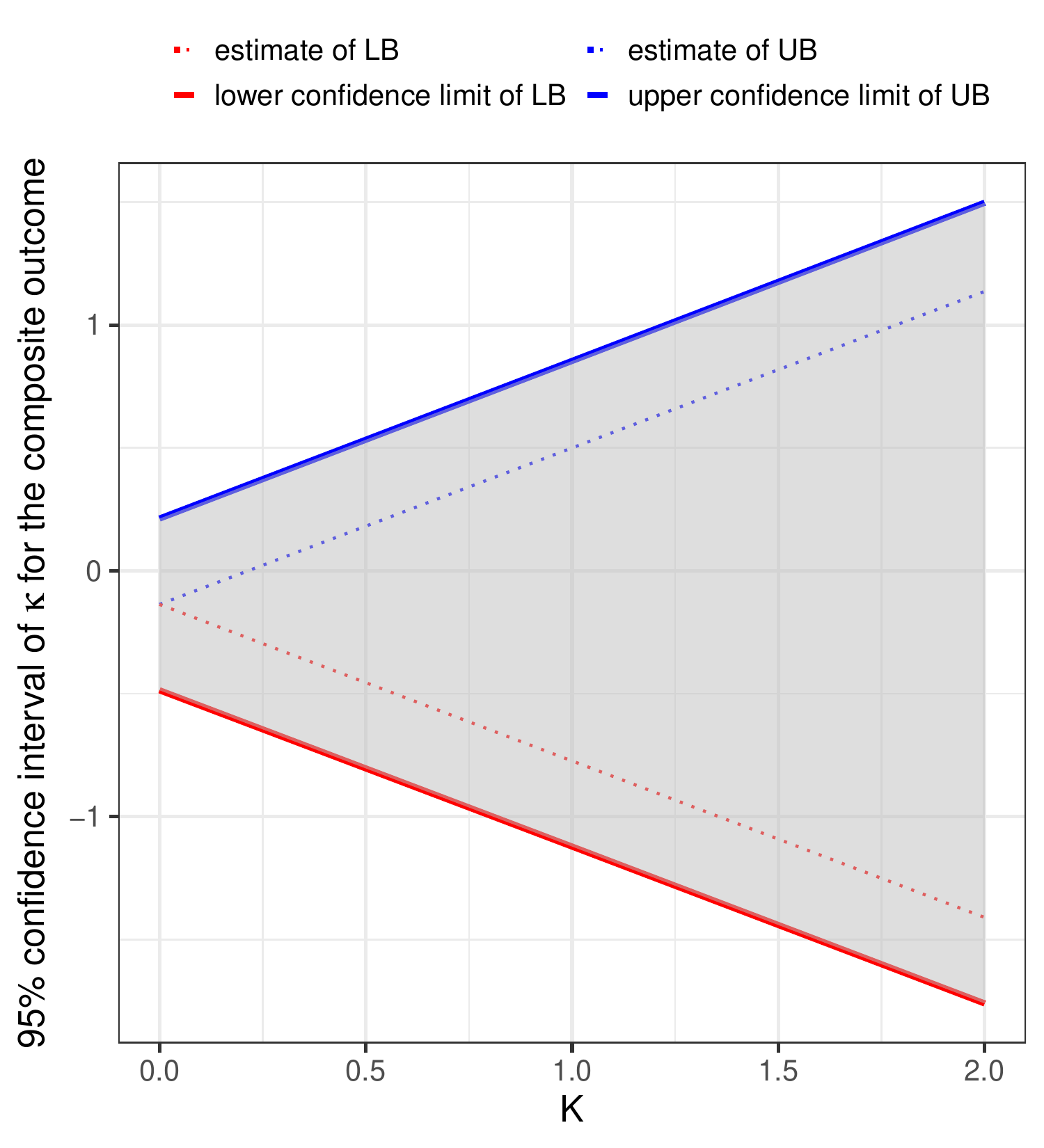} }}%
    \caption{ $95\%$ confidence interval plot of $\kappa$ for  (a) the  mortality rate with $K_0 = 0$ and $0 \leq K_1 \leq 0.03$ and (b) for the composite outcome with $0 \leq-K_0 =K_1 = K\leq 2$, under model $\mathcal{M}_\gamma$ with $\gamma = 0.0012.$ The dotted lines represent the point estimates while the solid lines represent the one-sided confidence limits of UB (in blue) and LB (in red), respectively.
    }%
    \label{fig:CI_plot}
\end{figure}

We next considered inferring the sample average treatment effect for the entire matched cohort $\textsf{M1}$ under $\mathcal{M}_\gamma$ with $\gamma = 0.0012$ based on Theorem \ref{thm: biased_inference}. The left panel of Figure \ref{fig:CI_plot} plots the 95\% confidence intervals of the mortality outcome for $81,766 \times 2 = 163,532$ participants in $\textsf{M1}$ against different hypothesized upper bound $K_1$ on the unidentified SATE among non-compliers. We assumed that compared to a low-level NICU, delivering at a high-level NICU would not hurt non-compliers so $K_0 = 0$. For instance, the 95\% confidence interval for the SATE was estimated to be $[0.25\%, 1.32\%]$ when $K_1 = 1\%$ and $[0.25\%, 1.97\%]$ when $K_1 = 2\%$. Results in Figure \ref{fig:CI_plot} simultaneously account for the sampling variability, the bias in the instrumental variable (controlled by $\mathcal{M}_\gamma$), and the unidentifiability of the SATE among non-compliers (controlled by the sensitivity parameters $K_0$ and $K_1$). Finally, the right panel of Figure \ref{fig:CI_plot} summarizes the results for the composite outcome for some selected $K_0$ and $K_1$ values.

Finally, we conducted a sensitivity analysis to report the sensitivity value \citep{zhao2019sensitivity} for the mortality outcome as discussed in Section \ref{subsec: inference under relaxed RI}. When $\gamma = 0.006$, the $95\%$ confidence interval for the effect ratio was estimated to be $[0.00, 2.91\%]$ and the lower confidence limit for the SATE also barely exceeded $0$ with $K_0 = 0$. 




\subsection{Subgroup analyses: Does the treatment effect vary by risk groups?}
\label{subsec: case study subgroup}

To further examine treatment effect heterogeneity, we performed an additional biased-randomization-based primary outcome analysis for the matched sample \textsf{M1} under Proposition \ref{prop: biased_inference for lambda} and model $\mathcal{M}_{\gamma}$ with $\gamma = 0.0012$ on four subgroups of mothers: (i) black mothers; (ii) non-black mothers who had reached 35 and whose gestational age were at most 30 weeks (non-black, high risk); (iii) non-black mothers who had not reached 35 and whose gestational age were at most 30 weeks (non-black, medium risk); and (iv) non-black mothers whose gestational age exceeded 30 weeks (non-black, low risk). Table \ref{table: primary analysis} summarizes the sample size, compliance rate, and outcomes of interest in the near and far groups for these subgroups in the matched sample $\textsf{M1}$.

\begin{figure}[H]%
    \centering
    \subfloat[\centering  Black mothers
    \label{fig:Death_CI_plot_blacks}]
{{\includegraphics[width=0.45\textwidth]{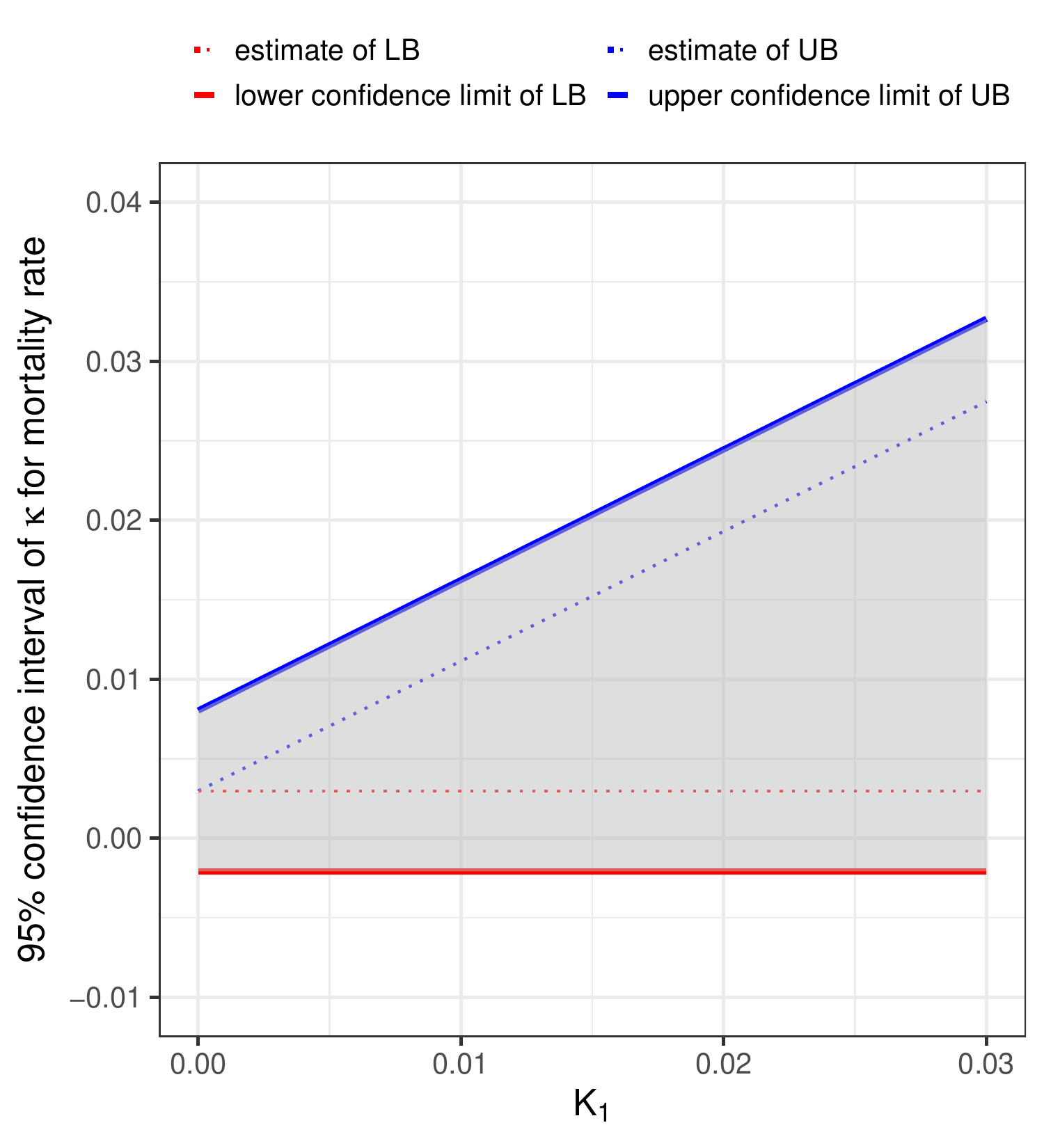} }}%
    \quad
    \subfloat[\centering Non-black, high-risk mothers
\label{fig:Death_CI_plot_nonblacks1}]{{\includegraphics[width=0.45\textwidth]{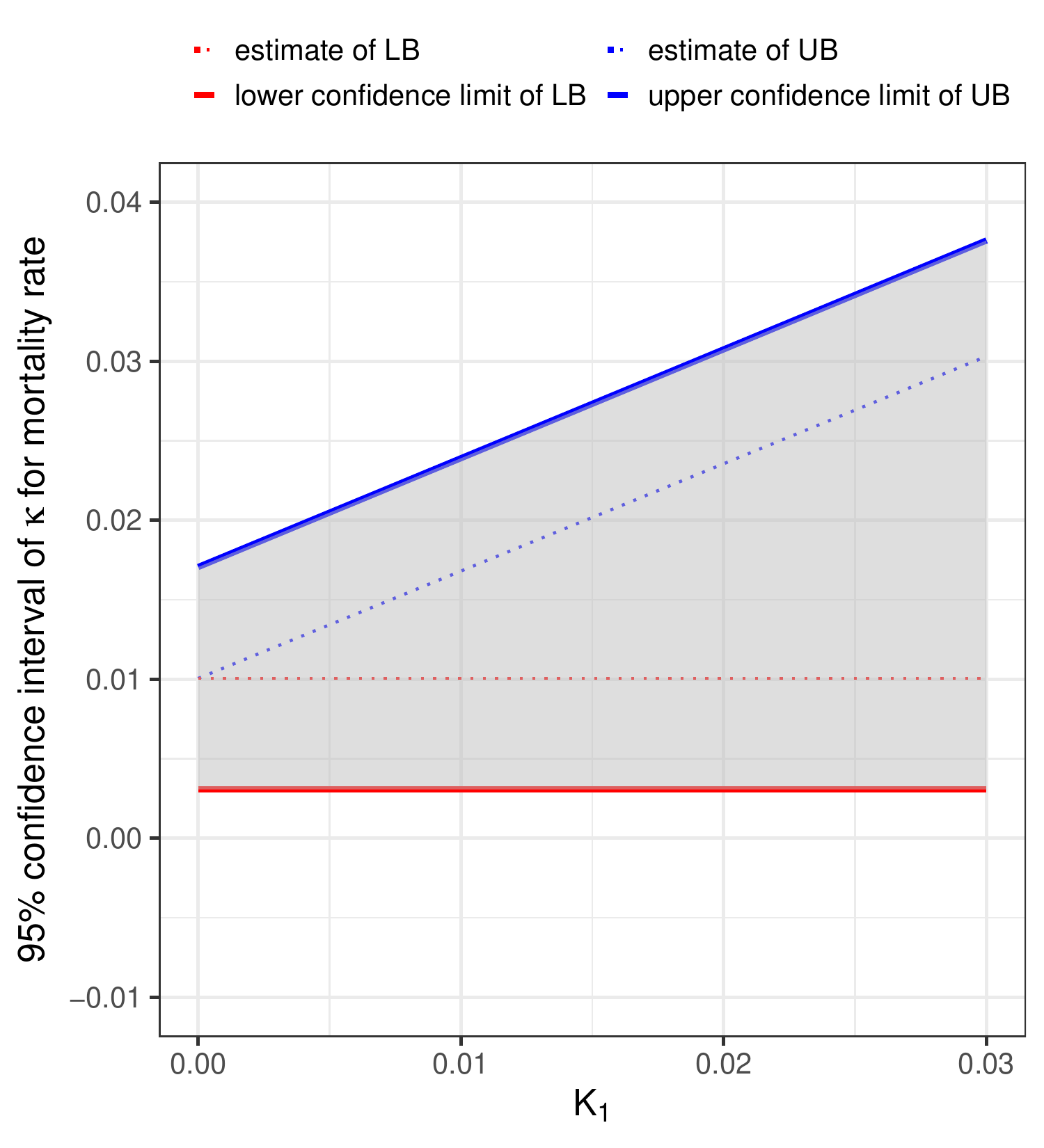} }}
    
\subfloat[\centering Non-black, medium-risk mothers
    \label{fig:Death_CI_plot_nonblacks2}]
{{\includegraphics[width=0.45\textwidth]{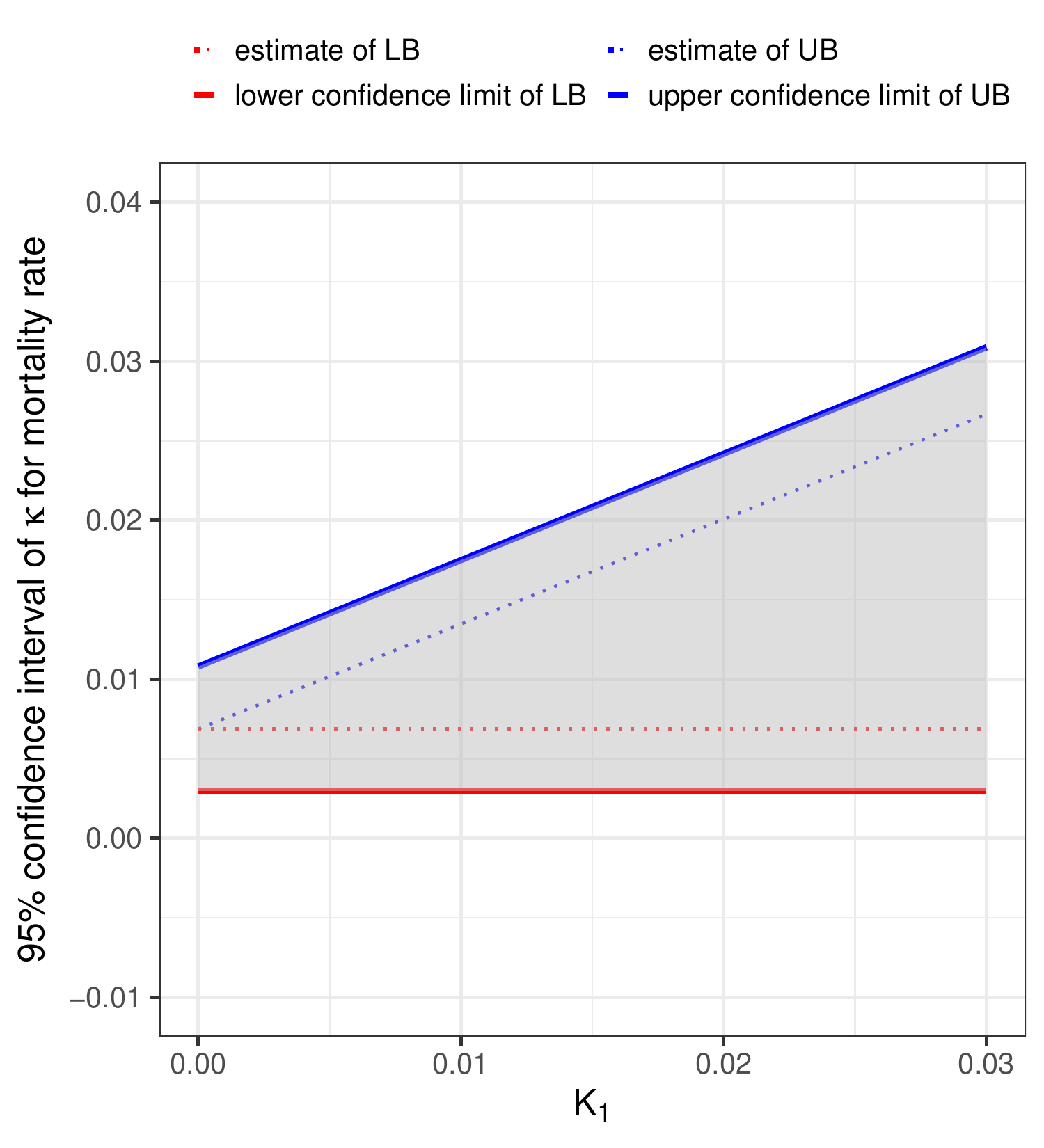} }}%
    \qquad
    \subfloat[\centering Non-black, low-risk mothers
\label{fig:Death_CI_plot_nonblacks3}]{{\includegraphics[width=0.45\textwidth]{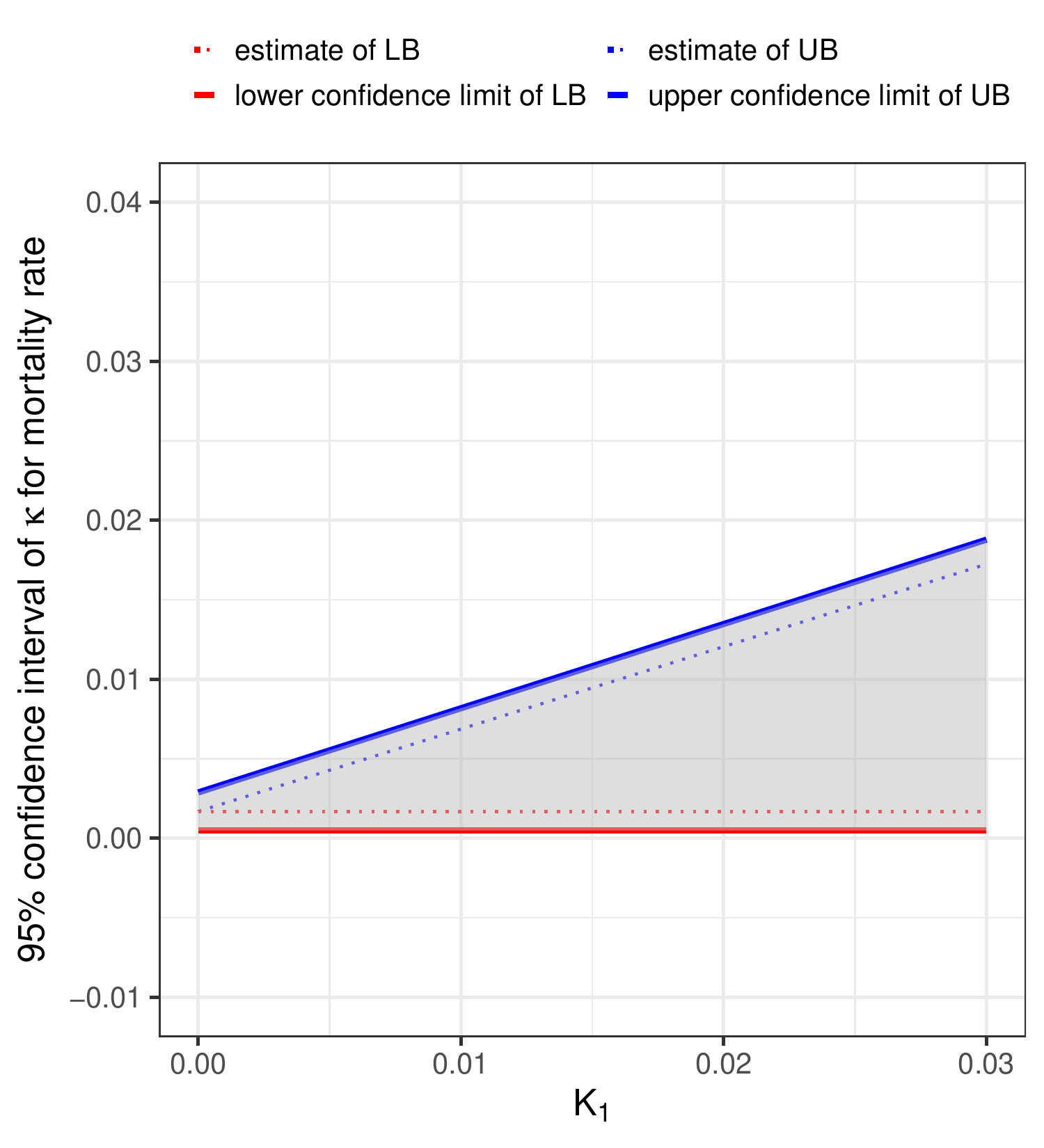} }}%
    \caption{ $95\%$ confidence interval plot of $\kappa$ for  (a) the  mortality rate with $K_0 = 0$ and $0 \leq K_1 \leq 0.03$ and (b) for the composite outcome with $0 \leq-K_0 =K_1 = K\leq 2$, under model $\mathcal{M}_\gamma$ with $\gamma = 0.0012.$ The dotted lines represent the point estimates while the solid lines represent the one-sided confidence limits of UB (in blue) and LB (in red), respectively. 
    }%
    \label{fig: CI_plot_mortality_subgroup}
\end{figure}

For the subgroup of black mothers, the estimated effect ratio (low-level NICU vs. high-level NICU) is $1.62\%$ with a $95\%$ confidence interval of $[-1.16\%, 4.50\%].$ Thus, we did not find evidence of a treatment effect among the $82.96\% - 64.64\% = 18.32\%$ compliers in $\textsf{M1}.$ Figure \ref{fig:Death_CI_plot_blacks} further plots the 95\% CIs for the sample average treatment effect among black mothers against different levels of $K_1$. Because of the low compliance rate among black mothers (even after strengthening) and a smaller sample size, both the effect ratio estimate and the partial identification bounds estimates are less precise compared to the analysis based on the entire study cohort in the design $\textsf{M1}$. The results from our analysis acknowledge that excess travel time yields rather limited information regarding a potential treatment effect of delivering at a high-level NICU vs. low-level NICU for black mothers in Pennsylvania, mostly due to the geography of the state. 

Among the non-black mothers, the effect ratio (low-level NICU vs. high-level NICU) was estimated to be $3.10\%$ with a $95\%$ confidence interval of $[0.92\%, 5.42\%]$ among the non-black, high-risk subgroup; $2.02\%$ with a $95\%$ confidence interval of $[0.85\%, 3.26\%]$ among the non-black, medium-risk subgroup; and $0.35\%$ with a $95\%$ confidence interval of $[0.10\%, 0.61\%]$ among the non-black, low-risk subgroup. Our analyses suggest that the clinical benefit of delivering at a high-level NICU appeared to be most pronounced among compliers in the high-risk subgroup, with a risk difference estimate almost 10-fold that among compliers in the low-risk subgroup.

Figure \ref{fig: CI_plot_mortality_subgroup} further plots the 95\% CIs of the sample average treatment effect for each non-black subgroup against different values of $K_1$. Comparing the partial identification bounds of three non-black risk groups, we have an impression that the low-risk subgroup appeared to minimally benefit from the high-level NICU; for instance, a $95\%$ CI for the SATE equals $[0.05\%, 0.39\%]$ when $K_1 = 0.2\%$, $[0.05\%, 0.50\%]$ when $K_1 = 0.4\%$, and $[0.05\%, 0.61\%]$ when $K_1 = 0.6\%$. Interestingly, among the four subgroups, the partial identification bounds are the narrowest for the non-black, low-risk subgroup, as a consequence of the large sample size ($n = 29,155 \times 2 = 58,310$) and high compliance rate ($65.9\% - 17.7\% = 48.2\%$) in this subgroup. 

Finally, we repeated the subgroup analysis for the composite outcome. We did not find any difference in the composite outcome among compliers in the black, high-risk non-black, and the medium-risk non-black subgroups; however, the effect ratio was estimated to be $-0.24$ with a $95\%$ confidence interval of $[-0.42, -0.07]$ for low-risk non-black mothers, providing some additional evidence that the triage system may consider sending low-risk mothers to low-level NICUs, especially under a resource constraint for high-level NICUs. For a $95\%$ confidence interval plot of the sample average treatment effect for each subgroup; see Supplemental Material C.

%% file: 8_Discussion.tex
\section{Discussion}
\label{sec: discussion}
Strengthening an instrumental variable when designing an observational study is an innovative approach that holds significant potential for enhancing empirical comparative effectiveness research. This technique can lead to important changes in the causal estimands and interpretations of derived treatment effects. In their discussion, \citet[Section 5]{baiocchi2010building} discussed aspects that would change as one switches between an unstrengthened IV design and a strengthened IV design, or between two different strengthened IV designs. Building on the insights from \citet{baiocchi2010building} and critiques from \citet{deaton2009instruments}, this article proposes strategies to harmonize various strengthened IV designs. One promising strategy is to complement the effect ratio estimate with partial identification bounds for the sample average treatment effect. While effect ratio estimates focus on compliers--a latent subgroup defined by the study design--partial identification bounds for the SATE provide insights applicable to the entire matched cohort, regardless of how the matched pairs or sets are formed. Strengthening the IV improves the compliance rate, which in turn enhances the precision of effect ratio estimates and leads to narrower partial identification bounds. This dual benefit underscores the value of strengthening IVs.

We investigated the impact of delivering in a high-level NICU compared to a low-level NICU using observational data from Pennsylvania and methodologies developed in this article, including an improved non-bipartite matching algorithm that emulates a randomized encouragement trial, as well as inferential methods for constructing valid level-$\alpha$ confidence intervals for partial identification bounds under randomization and IV-dependent, biased randomization schemes. Our analysis showed that delivering in a high-level NICU was associated with reduced mortality rates for preterm infants, with the effects varying significantly among different demographic groups. Notably, the most substantial benefits were observed for non-black, high-risk mothers, while the impact was marginal for non-black, low-risk mothers. This finding suggests that a triage system may not necessarily require transferring all mothers and their preterm infants to high-level NICUs. Interestingly, despite the elevated risk among black mothers and their potential for benefiting from high-level NICU care, our analysis did not find evidence of significant benefits in this group. This is likely attributed to the excess travel time being a weak IV for this subpopulation, even after efforts to strengthen it. 

Our work contributes to the debate between \citet{deaton2009instruments} and \citet{imbens2010better} regarding the internal and external validities of an IV analysis, whose tension becomes particularly pronounced when using a continuous IV. The SATE for the entire cohort is effectively a weighted average of the SATEs for various subpopulations (e.g., black mothers and non-black mothers categorized by risk levels). This perspective highlights a critical insight: if the IV serves as a weak encouragement for treatment uptake within a specific subpopulation, it yields limited information about that group. Consequently, despite our efforts to include black mothers in our representative matched cohort, our conclusion for this subpopulation remains inconclusive. Thus, our investigation sheds light on an important aspect of the IV validity debate: in the context of a continuous IV, striving for a more generalizable estimand may introduce greater uncertainty. Ultimately, we can only ascertain the treatment effect with precision for those subpopulations that respond strongly to the IV. 

For researchers, we recommend that the pursuit of a representative matched sample with a strengthened IV serve as the guiding principle in an IV-based study design, though achieving this goal depends heavily on available data and the many nuanced processes that shape it. Additionally, understanding which subpopulations respond to the IV is essential for planning the analysis effectively. Including subpopulations that do not respond to the IV should be optional albeit transparent to ensure that stakeholders understand the implications and limitations of the conclusions for such subpopulations.

In this study, we pre-specified several subgroups based on previous clinical findings \citep{hansen1986older,yang2014estimation,mathews2015infant}. With a binary treatment and in the context of testing Fisher's sharp null hypothesis, data-dependent methods that identify promising subgroups have been developed and deployed in matched observational studies \citep{hsu2013effect,hsu2015strong,lee2018discovering}. It is of great interest to develop methods that could identify subgroups most amenable to being strengthened and would potentially yield a large treatment effect. 